\documentclass[11pt,a4paper]{article}
\usepackage{jheppub}
\usepackage[utf8]{inputenc}
\usepackage{latexsym, graphicx} 
\usepackage{amsmath,amsfonts,amssymb}
\usepackage{mathrsfs}
\usepackage{tensor}

\providecommand{\Bt}{{\mathtt{B}}}
\renewcommand{\Bt}{{\mathtt{B}}}
\providecommand{\Ct}{{\mathtt{C}}}
\renewcommand{\Ct}{{\mathtt{C}}}
\providecommand{\Jt}{{\mathtt{J}}}
\renewcommand{\Jt}{{\mathtt{J}}}
\providecommand{\Ht}{{\mathtt{H}}}
\renewcommand{\Ht}{{\mathtt{H}}}

\providecommand{\Pt}{{\mathtt{P}}}
\renewcommand{\Pt}{{\mathtt{P}}}

\newcommand*{\scri}{\ensuremath{\mathscr{I}}}
\newcommand*{\lied}{\mathop{}\!\mathcal{L}}
\newcommand*{\dd}{\mathop{}\!d}

\newcommand*{\sA}{\sf{A}}
\newcommand*{\sB}{\sf{B}}
\newcommand*{\sC}{\sf{C}}

\newcommand*{\sa}{\sf{a}}

\title{The effective action of superrotation modes}
\author[a]{Kévin Nguyen,}
\author[b]{Jakob Salzer}
\emailAdd{kevin\_nguyen@g.harvard.edu}
\emailAdd{jsalzer@fas.harvard.edu}
\affiliation[a]{Black Hole Initiative, Harvard University, Cambridge, MA 02138, USA}
\affiliation[b]{Center for the Fundamental Laws of Nature, Harvard University, Cambridge, MA02138, USA}

\abstract{Starting from an analysis of four-dimensional asymptotically flat gravity in first order formulation, we show that superrotation reparametrization modes are governed by an Alekseev--Shatashvili action on the celestial sphere. This two-dimensional conformal theory describes spontaneous symmetry breaking of Virasoro superrotations together with the explicit symmetry breaking of more general Diff$(\mathcal{S}^2)$ superrotations. We arrive at this result by first reformulating the asymptotic field equations and symmetries of the radiative vacuum sector in terms of a Chern--Simons theory at null infinity, and subsequently performing a Hamiltonian reduction of this theory onto the celestial sphere.}

\begin{document}

\maketitle
\flushbottom

\section{Introduction}
The importance of asymptotic symmetries of gravitational and gauge theories has been appreciated to an ever increasing extent, starting with the discovery by Bondi, van der Burg, Metzner and Sachs (BMS)  of an infinite-dimensional symmetry group governing the phase space of General Relativity with asymptotically flat boundary conditions \cite{Bondi:1962px,Sachs:1962wk}. In particular, conservation laws associated to BMS symmetries --~also called \textit{supertranslations}~-- have been shown to yield Ward identities that are equivalent to Weinberg's soft graviton theorems and the displacement memory effect \cite{Strominger:2013jfa,Strominger:2014pwa,He:2014laa}; see the reviews \cite{Strominger:2017zoo,Compere:2018aar}.

In its original form, the BMS group contains supertranslations and Lorentz transformations where the latter act as $\textrm{Conf}(\mathcal{S}^2)$ global conformal transformations on the celestial sphere. More recently, the \textit{extended} BMS group \cite{Barnich:2010eb} and the \textit{generalized} BMS group \cite{Campiglia:2014yka,Campiglia:2015yka} have been considered two possible alternative extensions to the BMS group. Their new elements essentially act through Virasoro local conformal transformations and Diff($\mathcal{S}^2$) diffeomorphisms on the celestial sphere, respectively. The associated symmetry transformations go under the name of  \textit{superrotations} in both cases. Similarly to supertranslations, superrotations have been related to  subleading soft graviton theorems \cite{Cachazo:2014fwa,Campiglia:2014yka,Kapec:2014opa} and have been instrumental in the discovery of new memory effects \cite{Pasterski:2015tva,Compere:2018ylh}.

Supertranslations and superrotations are spontaneously broken, leading to an infinite degeneracy of gravitational vacua in asymptotically flat spacetimes \cite{Ashtekar:1981hw,Strominger:2013jfa}; an explicit construction of these vacua was presented in \cite{Compere:2016jwb,Compere:2018ylh,Adjei:2019tuj}. In the present work, we derive an effective action for superrotation modes on the celestial sphere which describes the explicit breaking of general superrotations down to the Virasoro subgroup, together with the spontaneous breaking of the latter. This is similar to the way the Schwarzian action describes the explicit breaking of $\textrm{Diff}(\mathcal{S}^1)$ down to $\textrm{SL}(2,\mathbb{R})$ in the context of gravity in two-dimensional anti-de Sitter (AdS) space \cite{Maldacena:2016upp}.

The approach employed in this work is reminiscent of other recent works on gravity in two and three dimensions. These lower-dimensional models exhibit no gravitational radiation so that all non-trivial states reside on the boundary. Consequently, it is possible to derive from the bulk theory an action that captures the dynamics of the boundary modes; see \cite{Coussaert:1995zp,Cotler:2018zff,Cotler:2019nbi,Merbis:2019wgk} for constructions on $\textrm{AdS}_3,\textrm{dS}_3$, and three-dimensional flat space, respectively. In the case of pure three-dimensional flat space, one finds that the boundary action at null infinity further reduces to an action on its (one-dimensional) boundary. More precisely, the analysis of \cite{Carlip:2016lnw} shows that the action obtained in this way is a Schwarzian action for superrotation modes. As will be seen, the action we obtain in the present work is the natural generalization of this result to four-dimensional asymptotically flat gravity.

In the remainder of this introduction we present an outline of our work and highlight some of the main steps along the way.

In section~\ref{sec:section2} we provide a description of four-dimensional asymptotically flat gravity in a first order formulation where the independent variables are frame fields and associated spin connection coefficients. Imposing the standard Newman--Unti boundary conditions at future null infinity $\scri^+$ \cite{Newman:1962cia}, we show that the set of residual large gauge symmetries generates the extended/generalized $\mathfrak{bms_4}$ algebra together with boundary Weyl rescalings, in agreement with an earlier analysis performed by Barnich and Lambert \cite{Barnich:2011ty}. The formalism which we use closely follows that of Korovin \cite{Korovin:2017xqu} and differs only in the use of the more restrictive Newman--Unti boundary conditions.

We proceed in section~\ref{sec:radiative-vacua} with an asymptotic analysis of the field equations, restricting our attention to field configurations without gravitational radiation going through~$\scri^+$. These \textit{radiative vacua} differ asymptotically from Minkowski space by large gauge transformations \cite{Compere:2016jwb,Compere:2018ylh}. Hence, radiative vacua spontaneously break the asymptotic symmetry group. This spontaneous breaking is characterized by \textit{supertranslation} and \textit{superrotation} modes whose description is the subject of the present work. As a first step in that direction, we organize the field equations into successive layers that may be solved iteratively near $\scri^+$. Interestingly, the first layer precisely coincides with the equations of motion of an $\mathfrak{so}(3,1)=\mathfrak{conf}(2)$ Chern-Simons theory defined at null infinity,
\begin{equation}
\label{eq:CSaction-intro}
  S=\frac{k}{4\pi}\int_{\scri^+} \langle A\wedge \dd A+\frac{2}{3}A\wedge A\wedge A\rangle\,,
\end{equation}
where the algebra-valued gauge connection $A$ may be decomposed as
\begin{equation}
\label{eq:CSconnection-intro}
 A=h\Ht+e^{\sa}\Pt_{\sa}+b^{\sa}\Bt_{\sa}+\omega \Jt\,.
\end{equation}
With hindsight, the appearance of the Lorentz group SO(3,1) is barely surprising as it is the symmetry group of the light-cone of which null infinity is a close cousin. 

Except for its component $b^{\sa}$, the gauge field of the above Chern--Simons theory admits a natural interpretation in terms of the intrinsic conformal geometry\footnote{We also refer the reader to the recent work \cite{Herfray:2020rvq} for a  mathematical description of null infinity from an intrinsic conformal geometric perspective. } at null infinity~\cite{Gryb:2012qt}. In order to describe the asymptotic structure of a four-dimensional spacetime $\mathcal{M}$ equipped with metric $g_{\alpha \beta}$, it is convenient to introduce its conformal compactification \cite{Penrose:1965am,Geroch1977}. To this end, one introduces the rescaled \emph{unphysical metric}
\begin{equation}
  \label{eq:compactification}
  \tilde{g}_{\alpha\beta}=\Omega^2g_{\alpha\beta}\,,
\end{equation}
where the conformal factor $\Omega$ is chosen in such a way that $\tilde{g}_{\alpha\beta}$ is everywhere finite. Null infinity $\scri$ is then defined as the locus where this conformal factor vanishes, and is understood as the spacetime conformal boundary. When pulled back to this boundary surface $\Omega=0$, the unphysical metric $\tilde{g}$ induces a degenerate boundary metric $q$ whose kernel is spanned by the vector $n$ normal to $\scri$. This defines a Carrollian structure on $\scri$. The components $e^{\sa}$ of the Chern--Simons connection play the role of frame fields for this degenerate metric,
\begin{equation}
\label{eq:qzweibein}
q_{\mu\nu}=\delta_{\sf{ab}} e^{\sf{a}}_\mu e^{\sf{b}}_\nu, \qquad e^{\sa}\cdot n=0\,.
\end{equation}
The choice of conformal factor $\Omega$ in \eqref{eq:compactification} is not unique, however. Any change $\Omega\mapsto e^\eta \Omega$ still yields a regular unphysical metric, and induces a Weyl rescaling $q\mapsto e^{2 \eta} q$ at $\scri$. Obviously, this freedom should be considered unphysical or pure gauge, while any sensible physical quantity computed at $\scri$ ought to be Weyl-invariant. One might therefore find convenient to work with a torsion-free \textit{Weyl covariant derivative} $\hat{D}$ at $\scri$ which, in contrast to more conventional metric compatible derivative operators, is constructed such as to be invariant under Weyl rescalings \cite{Weyl:1918ib}. We review this construction in appendix~\ref{app:Weyl-connection}. A Weyl covariant derivative satisfies
\begin{equation}
\label{nonmetricity}
\hat{D}_\rho q_{\mu\nu}=h_\rho q_{\mu\nu}\,,
\end{equation}
where the \textit{Weyl vector} $h$ is a priori arbitrary and transforms like a gauge field for Weyl rescalings, $h \mapsto h+d\eta$. As it turns out, the component $h$ of the Chern--Simons gauge field \eqref{eq:CSconnection-intro} precisely behaves as a Weyl vector. The spin connection $\hat{\omega}$ associated to the Weyl derivative $\hat{D}$ satisfies
\begin{equation}
\label{eq:omegahat}
\hat{\omega}\indices{^{\sf{ab}}}=\omega^{\sf{ab}}-h \delta^{\sf{ab}}\,,
\end{equation}
where $\omega^{\sf{ab}}$ is a metric compatible spin connection with torsion, and the quantity $\omega$ appearing in \eqref{eq:CSconnection-intro} is its Hodge dual,
\begin{equation}
\omega^{\sf{ab}}=-\omega \epsilon^{\sf{ab}}\,.
\end{equation}

The main surprise comes from the field $b^{\sa}$ which does not admit an intrinsic geometric interpretation at null infinity. Instead we identify it with the pull-back to $\scri$ of the Schouten tensor $\tilde{S}_{\alpha\beta}$ associated to the unphysical bulk metric $\tilde{g}_{\alpha \beta}$,
\begin{equation}
\label{eq:Schouten}
S_{\mu\nu}\equiv\tilde{S}_{\mu\nu}\big|_{\scri}= -2b^{\sa}_\mu e^{\sa}_\nu, \qquad \tilde{S}_{\alpha \beta}=\tilde{R}_{\alpha \beta}-\frac{1}{6} \tilde{R}\ \tilde{g}_{\alpha \beta}.
\end{equation}
This quantity encodes information about \textit{gravitational radiation} passing through $\scri$, but is not invariant under boundary Weyl rescalings. The News tensor $N_{\mu\nu}$ contains the physical Weyl-invariant information,
\begin{equation}
N_{\mu\nu}\equiv S_{\mu\nu}-\rho_{\mu\nu}\,,
\end{equation}
where the Geroch tensor $\rho_{\mu\nu}$ is a geometric tensor intrinsically defined on $\scri$ whose transformation under Weyl rescalings precisely cancels that of $S_{\mu\nu}$ \cite{Geroch1977}. As already mentioned, solutions of the Chern--Simons theory \eqref{eq:CSaction-intro} correspond to radiative vacua of asymptotically flat gravity. Insisting on having vanishing News associated to such radiative vacua\footnote{\label{footnote:News}Sometimes a different convention has been adopted where $\rho_{\mu\nu}$ is fixed once and for all, such that the various vacuum configurations $\rho_{\mu\nu}=S_{\mu\nu}^{\text{vac}}$ translate into various vacuum News configurations $N_{\mu\nu}^{\text{vac}}$ instead \cite{Compere:2016jwb,Compere:2018ylh}.}, the Chern--Simons theory therefore describes the corresponding set of tensors $\rho_{\mu\nu}$ that characterize the respective vacua.

Radiative vacua and their associated field configurations have been described in \cite{Compere:2016jwb,Compere:2018ylh}. In particular, it has been shown that they are all related by superrotation symmetries, i.e.~by mappings of the \textit{celestial sphere} onto itself \cite{Adjei:2019tuj}. Using complex stereographic coordinates $(z,\bar{z})$, superrotations are thus generated by arbitrary reparametrizations
\begin{equation}
z'=\Pi(z,\bar{z}), \qquad \bar{z}'=\bar{\Pi}(z,\bar{z})\,.
\end{equation}
In section~\ref{sec:reduct-bound-theory} we proceed with the Hamiltonian reduction of the Chern--Simons theory and derive an action for this reparametrization mode,
\begin{equation}
  \label{eq:finalintro}
  S\left[\Pi\right]=\frac{t}{16\pi}\int_{\mathcal{S}^2}\dd^2 z\
  \frac{\partial_ z \partial_{\bar{z}}\Pi\ \partial_z^2\Pi}{(\partial_z \Pi)^2}+\frac{\bar{t}}{16\pi}\int_{\mathcal{S}^2}\dd^2 z \
  \frac{\partial_{\bar{z}}\partial_z\bar{\Pi}\ \partial_{\bar{z}}^2\bar{\Pi}}{(\partial_{\bar{z}} \bar{\Pi})^2}\,.
\end{equation}
It is recognized as a complex version of the geometric action on a coadjoint orbit of the Virasoro group put forward by Alekseev and Shatashvili \cite{Alekseev:1988ce}.  The derivation of this conformal field theory for the superrotation reparametrization mode is the main result of this work. The above effective action offers a unified treatment of Virasoro and Diff($\mathcal{S}^2$) superrotations and at the same time highlights the difference between them. Indeed, Diff($\mathcal{S}^2$) superrotations $\Pi(z, \bar{z})$ have non-zero action and are therefore explicitly broken. On the other hand, holomorphic Virasoro transformations $\Pi(z)$ have zero action and may be understood as labeling spontaneously broken \textit{superrotation vacua}.

\paragraph{Conventions.} 
Indices $\alpha,\beta,\gamma,...$ denote four-dimensional, $\mu,\nu,...$ denote three-dimensional, and $i,j,k,..$ denote two-dimensional coordinate indices. Similarly, sans serif letters $\sA,\sB,\sC,...$ denote four-dimensional frame indices, while $\sf{a},\sf{b},\sf{c},...$ denote two-dimensional frame indices. For the latter, which are raised and lowered with $\delta_{\sf{ab}}$, we will not distinguish between covariant and contravariant indices.

\section{Asymptotically flat spacetimes}
\label{sec:section2}
In this section we start by describing the Newman--Unti gauge and associated boundary conditions, both in metric and first order formulation. We then discuss residual asymptotic symmetries, recovering the extended/generalized BMS$_4$ symmetries together with boundary Weyl rescalings \cite{Barnich:2011ty}.

\subsection{Newman--Unti gauge}
\label{sec:newman-unti-solution}
For definiteness, we restrict the asymptotic analysis to the neighborhood of future null infinity $\scri^+$, but a similar analysis may be performed near past null infinity $\scri^-$.
Our starting point is the metric in \emph{Newman--Unti (NU) gauge} \cite{Newman:1962cia}, characterized by
\begin{align}
  \label{eq:NUgf}
g_{ur}=-1, \qquad g_{rr}=g_{ri}=0\,,
\end{align}
such that the most general metric in this gauge is of the form
\begin{equation}
  \label{eq:NUmetric}
  \dd s^2=g_{uu} \dd u^2-2\dd r \dd u+g_{ui} \dd u \dd x^i+g_{ij}\dd x^i\dd x^j,
\end{equation}
with falloff conditions
\begin{subequations}
\label{eq:falloff-metric}
\begin{align}
g_{uu}&=O(r),\\
g_{ui}&=O(r^0),\\
  \label{eq:19}
  g_{ij}&=r^2 \gamma_{ij} +r C_{ij}+O(r^0).
\end{align}
\end{subequations}
This choice of gauge is reached by demanding that slices of constant $u$ be lightlike and that their normal vector, lying in the lightlike hypersurface, generate geodesics with affine parameter $r$. The remaining coordinates $x^i$ label geodesics in a constant $u$-hypersurface. The freedom in scaling $r$ is used to set $g_{ur}=-1$, while its origin is fixed such as to enforce the additional condition of having no term of order $r^{-2}$ in
\begin{equation}
  \label{eq:16}
  \rho\equiv -\frac{1}{4} g^{ij}\partial_rg_{ij}=-r^{-1}+O(r^{-3}).
\end{equation}
In particular, tracelessness of the \textit{shear tensor} follows from this condition,
\begin{equation}
  \label{eq:Ctraceless}
  \gamma^{ij}C_{ij}=0\,.
\end{equation}
As an aside, let us mention that the more widely used \emph{Bondi gauge}, defined by the gauge fixing conditions
\begin{equation}
  \label{eq:Bondigf}
g_{\bar{r}\bar{r}}=g_{\bar{r}i}=0, \qquad \partial_{\bar{r}} \det \left(\frac{ g_{ij}}{\bar{r}^{4}}\right)=0\,,
\end{equation}
differs from the NU gauge only by the choice of radial coordinate $\bar{r}$. The relation between these two coordinate systems is simply given by  \cite{Kroon:1998dv,Barnich:2011ty}
\begin{equation}
  \label{eq:BNUr}
\bar{r}=\left(\frac{\det g_{ij}}{\det \gamma_{ij}}\right)^{\frac{1}{4}}=r+O(r^{-1})\,.
\end{equation}
The difference between these two gauges appears only at subleading order and, being a choice of gauge, should not affect the computation of physical quantities such as surface charges \cite{Barnich:2011ty}. We find that the subsequent analysis is most transparent and natural in NU gauge.

We turn to a description of the above metric in terms of frame fields which, together with the spin connection, are the independent fields in a first order formulation of gravity. The frame fields $E_\alpha^{\sA}$, which form a vector basis in the spacetime tangent bundle, satisfy
\begin{equation}
  \label{eq:gto}
  g_{\alpha\beta}=\eta_{\sA\sB}E_\alpha^{\sA}E_\beta^{\sB}\,.
\end{equation}
We will label the four covectors by $\sf{\hat{u},\hat{r},a}$ where $\sf{a}=1,2$. The Lorentzian tangent space metric is taken to be
\begin{equation}
  \label{eq:tmetric}
\eta_{\sf{\hat{u}}\hat{r}}=-1\,, \qquad \eta_{\sa\sf{b}}=\delta_{\sa\sf{b}}\,.
\end{equation}
In particular, there is no difference between a lower or upper frame index $\sa$. The spin connection $\Omega\indices{_\alpha^{\sA}_{\sB}}$ contains the Ricci rotation coefficients in this frame basis,
\begin{equation}
  \label{eq:spincon}
\nabla_{\alpha} E^{\sA}_\beta=-\Omega\indices{_\alpha^{\sA}_{\sB}} E^{\sB}_\beta\,.
\end{equation}
Metric compatibility of the covariant derivative $\nabla$ requires antisymmetry of the spin connection with respect to frame indices, $\Omega^{[\sA\sB]}=0$, which we assume for consistency with the metric formulation of General Relativity. In a first order formulation thereof, the spin connection is considered an independent field. It can be completely solved in terms of the frame fields through its own equation of motion, the vanishing torsion constraint
\begin{equation}
\label{eq:torsion}
T^{\sA}=0\,, \qquad T^{\sA}\equiv dE^{\sA}+\Omega\indices{^{\sA}_{\sB}} \wedge E^{\sB}\,.
\end{equation}
The covariant derivative $\nabla$ then reduces to the standard metric compatible and torsion-free Levi-Civit\'a one.

In the first order formulation one has the freedom to perform general coordinate transformations, as well as local Lorentz transformations (LLT) that leave the tangent space metric $\eta_{\sA\sB}$ invariant. We use this freedom to impose the gauge conditions
\begin{equation}
\label{eq:NU1}
E^{\sf{\hat{u}}}_u=E^{\sf{\hat{r}}}_r=1, \qquad  E^{\sf{\hat{u}}}_r=E^{\sf{\hat{u}}}_i=E^{\sf{a}}_r=0, \qquad \Omega\indices{_r^{\sa\sf{b}}}=\Omega\indices{_r^{\hat{r}\sa}}=0\,.
\end{equation}
Vanishing of the torsion components $T^{\sf{\hat{u}}}_{ru}$ and $T^{\sf{\hat{u}}}_{ri}$ directly yields
\begin{equation}
  \label{eq:NU2}
  \Omega\indices{_r^{\sA\sB}}=0, \qquad \text{for all} \quad \sA, \sB\,,
\end{equation}
such that we may consider these `extended' conditions for simplicity. In section~\ref{sec:residual-sym}, we will show that the above gauge conditions leave some \textit{residual} gauge symmetries, to be interpreted as asymptotic symmetries. Thus, the frame fields take the form
\begin{subequations}
\begin{align}
E^{\sf{\hat{u}}}&=\dd u\,,\\
E^{\sf{\hat{r}}}&=\dd r+E^{\sf{\hat{r}}}_u \dd u+E^{\sf{\hat{r}}}_i \dd x^i\,,\\
E^{\sf{a}}&=E^{\sf{a}}_u \dd u+E^{\sf{a}}_j \dd x^j\,.
\end{align}
\end{subequations}
The corresponding dual vector fields are given by
\begin{subequations}
\begin{align}
E_{\sf{\hat{u}}}&=\partial_u+\left(E^{\sf{\hat{r}}}_i E^i_{\sf{a}} E^{\sf{a}}_u - E^{\sf{\hat{r}}}_u \right)\partial_r-E^{\sf{a}}_u E^i_{\sf{a}} \partial_i\,,\\
E_{\sf{\hat{r}}}&= \partial_r\,,\\
E_{\sf{a}}&=E_{\sf{a}}^i \partial_i-E^i_{\sf{a}} E^{\sf{\hat{r}}}_i \partial_r\,,
\end{align}
\end{subequations}
where $E_{\sa}^i$ is defined to be the inverse of $E^{\sa}_i$ and therefore satisfies
\begin{equation}
E_{\sa}^i E^{\sa}_j=\delta^i_j\,, \qquad E_{\sa}^i E^{\sf{b}}_i=\delta_{\sa}^{\sf{b}}\,.
\end{equation}
Taken together, the gauge-fixing conditions \eqref{eq:NU1} and
\eqref{eq:NU2} define the NU gauge in terms of frame fields and spin
connection \cite{Newman:1962cia}. Since the NU gauge is usually
discussed in terms of Newman--Penrose (NP) quantities, we provide an
explicit translation of our gauge-fixing conditions to the NP
formulation in appendix \ref{sec:comp-newm-penr}.

In order to satisfy the NU fall-off conditions \eqref{eq:falloff-metric}, we require
\begin{subequations}
\label{eq:framefall}
\begin{align}
E^{\sf{\hat{r}}}_\mu&=r h_\mu+h^{(0)}_\mu +O(r^{-1}),\\
E^{\sf{a}}_\mu&=r e^{\sf{a}}_\mu+e^{(0)\sf{a}}_\mu+O(r^{-1}),
\end{align}
\end{subequations}
where all asymptotic fields depend on $u,x^i$, and with 
\begin{equation}
\label{eq:NU-bc}
e^{\sa}_u=0, \qquad e^{(0)\sa}_u=e_{\sa}^i h_i\equiv h_{\sa}.
\end{equation}
The field $e^{\sa}_\mu$ plays the role of frame field for the three-dimensional degenerate metric $q_{\mu\nu}$ on $\scri^+$, whose pull-back onto two-dimensional spatial sections thereof is $\gamma_{ij}$. We will refer to any such spatial section as the \textit{celestial sphere}. It is therefore natural to apply $e^{\sa}_i$ to fields `living' at $\scri^+$ -- fields appearing in the asymptotic expansion \eqref{eq:framefall} and which have no dependence on the radial coordinate -- in order to switch between coordinate indices~$i$ and frame indices~$\sa$.
 We may similarly decompose spatial vectors onto this frame field basis. In particular, we write
\begin{align}
e^{(0)\sa}_i=\frac{1}{2} C_{\sa \sf{b}} e^{\sf{b}}_i.
\end{align} 
Upon imposing the vanishing torsion constraint \eqref{eq:torsion}, it may be shown that the gauge condition $\Omega\indices{_r^{\sf{ab}}}=0$ yields
\begin{equation}
  \label{eq:Csym}
  C_{[\sa \sf{b}]}=0.
\end{equation} 

As mentioned in the introduction and explained in more details in section \ref{sec:radiative-vacua} and appendix \ref{app:Weyl-connection}, the frame field $h_\mu$ appearing in \eqref{eq:framefall} is interpreted as Weyl vector for the connection induced at the boundary. Since it is pure gauge $h_\mu$, it must be fixed on all of $\scri^+$ in order to solve the field equations.

We further restrict the phase space to field configurations satisfying the boundary condition
\begin{equation}
\label{eq:hu}
e^{\sa}_i(u,x)=\Theta(u,x) \bar{e}^{\sa}_i(x), \qquad h_u=\partial_u \ln \Theta\,.
\end{equation}
In order to make contact with recent works \cite{Campiglia:2014yka,Campiglia:2015yka,Compere:2018ylh}, we assume that the field $\bar{e}^a_i$ is completely arbitrary and only its determinant is fixed to yield the volume form of the unit round metric on the celestial sphere. As we will see explicitly in the next section, these boundary conditions imply that the asymptotic symmetry algebra contains superrotations. The field $\Theta$ in \eqref{eq:hu} on the other hand can be regarded as the ambiguity in the conformal factor used to define the unphysical metric \eqref{eq:compactification}. It is pure gauge and can thus be fixed to any desired value.

The freedom in choosing the conformal factor is often used to go to a \textit{Bondi conformal frame} in which
\begin{equation}
  \label{eq:81}
  \tilde{\nabla}_\alpha\tilde{\nabla}_\beta\Omega|_{\scri^+}=0\,
\end{equation}
holds
where $\tilde{\nabla}$ is the Levi-Civitá connection associated to $\tilde{g}$. In the present context this corresponds to setting $h_u=0$. While the use of a Bondi conformal frame is convenient for discussing quantities defined strictly on $\scri$, it is less so if one is interested in the behavior of fields in a neighborhood of spatial infinity. In fact, using the Ashtekar--Hansen definition of asymptotic flatness near null and spatial infinity, one finds that a conformal factor obeying \eqref{eq:81} is incompatible with asymptotic flatness at spatial infinity \cite{Ashtekar:1978zz, Wald:1984gr}. Since we are ultimately interested in fields defined at $\scri
^+_-$, we will refrain from setting $h_u=0$ in general.
Nevertheless, in order to compare with results in the literature we will sometimes go to a Bondi conformal frame by which we mean in a slight abuse of terminology the stronger condition
\begin{equation}
  \label{eq:Bondimisuse}
h_\mu=0\,
\end{equation}
We are not aware of any pre-existing treatment of the less restrictive gauge $h_i\neq 0$.

In terms of frame field components, the asymptotic expansion of the metric \eqref{eq:NUmetric} is
\begin{subequations}
\label{eq:asexp}
\begin{align}
g_{uu}&=-2r h_u+(-2h_u^{(0)}+h^{\sf{a}} h_{\sf{a}})+O(r^{-1})\,,\\
g_{ui}&=(h_{\sa}e^{(0)\sf{a}}_i-h_i^{(0)}+e^{\sf{a}}_ie^{(1)\sf{a}}_u)+O(r^{-1})\,,\\
g_{ij}&=r^2\gamma_{ij}+r C_{ij}+O(r^{0})\,,
\end{align}
\end{subequations}
with
\begin{equation}
\gamma_{ij}=\delta_{\sa \sf{b}} e^{\sf{a}}_i e^{\sf{b}}_j, \qquad C_{ij}=e^{\sa}_i e^{\sf{b}}_j C_{\sa \sf{b}}\,.
\end{equation}
As in the metric formulation, we require tracelessness of the shear tensor $C_{ij}$ as an additional gauge-fixing condition,
\begin{equation}
  \label{eq:gfC}
\gamma^{ij}C_{ij}=\delta^{\sa \sf{b}} C_{\sa \sf{b}}=0\,.
\end{equation}
In the first order formulation of gravity, we also need to provide fall-off conditions for the spin connection. As stated above, the latter will be uniquely determined through the vanishing torsion constraint \eqref{eq:torsion}. Fall-off conditions that are compatible with this unique solution take the form
\begin{subequations}
  \label{eq:spinex}
\begin{align}
\Omega\indices{_\mu^{\sf{\hat{u}}\hat{r}}}&=\omega\indices{_\mu^{\sf{\hat{u}}\hat{r}}}+O(r^{-2})\,,\\
\Omega\indices{_\mu^{\sf{\hat{u}}\sa}}&=\omega\indices{_\mu^{\sf{\hat{u}}\sa}}+O(r^{-2}),\\
\Omega\indices{_\mu^{\sf{\hat{r}}\sa}}&=b^{\sa}_\mu +r^{-1}\omega_\mu^{(1)\sf{\hat{r}}\sa}+O(r^{-2}),\\
\Omega\indices{_\mu^{\sf{ab}}}&=\omega\indices{_\mu^{\sf{ab}}}+O(r^{-2})\,.
\end{align}
\end{subequations}

The boundary conditions defined above agree with those originally
defined in \cite{Newman:1962cia} and more recently discussed in
\cite{Barnich:2011ty,Barnich:2016lyg,Barnich:2019vzx}. The field $h_i$ is usually eliminated by a null rotation around
$E^{\sf{\hat{u}}}$. Refraining from doing this will allow us to identify a gauged $\mathfrak{so}(3,1)=\mathfrak{conf}(2)$ symmetry algebra governing the asymptotic fields.

\subsection{Residual asymptotic symmetries}
\label{sec:residual-sym}
We now discuss residual gauge symmetries that preserve the NU gauge-fixing conditions \eqref{eq:NU1}. In the first order formulation of gravity, gauge symmetries include covariant general coordinate transformations\footnote{Covariant general coordinate transformations preserve the tensorial nature of fields with respect to internal gauge symmetries, such as LLT in this case. They may be thought of as standard diffeomorphisms corrected by simultaneous internal gauge transformations. See Chapter~11 of \cite{Freedman:2012zz}.} (cgct) as well as local Lorentz transformations (LLT). The infinitesimal transformation of the frame fields and the spin connection is given by 
\begin{subequations}
\begin{align}
\delta E^{\sA}_\alpha&=\xi^\beta \partial_\beta E^{\sA}_\alpha+E^{\sA}_\beta \partial_\alpha \xi^\beta-\bar{\Lambda}\indices{^{\sA}_{\sB}} E^{\sB}_\alpha\,,\\
 \delta \Omega\indices{_\alpha^{\sA\sB}}&=\xi^\beta \partial_\beta \Omega\indices{_\alpha^{\sA\sB}}+\Omega\indices{_\beta^{\sA\sB}}\partial_\alpha \xi^\beta+\partial_\alpha \bar{\Lambda}^{\sA\sB}+\Omega\indices{_\alpha^{\sC\sA}} \bar{\Lambda}\indices{^{\sB}_{\sC}}-\Omega\indices{_\alpha^{\sC\sB}} \bar{\Lambda}\indices{^{\sA}_{\sC}}\,,
\end{align}
\end{subequations}
where 
\begin{equation}
\bar{\Lambda}\indices{^{\sA}_{\sB}}\equiv \Lambda\indices{^{\sA}_{\sB}}-\xi^\alpha \Omega\indices{_\alpha^{\sA}_{\sB}}\,.
\end{equation}
The vector field $\xi$ generates cgct while the antisymmetric matrix $\Lambda\indices{^{\sA\sB}}$ generates LLT. These infinitesimal transformations may be written alternatively as \cite{Freedman:2012zz,Korovin:2017xqu}
\begin{subequations}
\label{eq:gauge-transf-2}
\begin{align}
\delta E^{\sA}_\alpha&=\partial_\alpha \xi^{\sA}-\Lambda\indices{^{\sA}_{\sB}} E^{\sB}_\alpha+\xi^{\sB} \Omega\indices{_\alpha^{\sA}_{\sB}}+\xi^\beta T^{\sA}_{\beta\alpha}\,,\\
\delta \Omega\indices{_\alpha^{\sA\sB}}&=\partial_\alpha \Lambda^{\sA\sB}+\Omega\indices{_\alpha^{\sC\sA}} \Lambda\indices{^{\sB}_{\sC}}-\Omega\indices{_\alpha^{\sC\sB}} \Lambda\indices{^{\sA}_{\sC}}+\xi^\beta R_{\beta\alpha}^{\sA\sB}\,,
\end{align}
\end{subequations}
where $\xi^{\sA}=\xi^{\alpha}E^{\sA}_\alpha$ and the curvature tensor has been introduced,
\begin{equation}
R^{\sA\sB}=\dd \Omega^{\sA\sB}+\Omega\indices{^{\sA}_{\sC}} \wedge \Omega^{\sC\sB}\,.
\end{equation}
Preservation of the gauge-fixing conditions \eqref{eq:NU1} impose restrictions on the symmetry generators $\xi$ and $\Lambda^{\sA\sB}$. Looking first at transformations of the frame fields, we obtain the constraints
\begin{subequations}
\begin{align}
0&=\partial_r \xi^u\,,\\
0&=\partial_u \xi^u+\bar{\Lambda}^{\sf{\hat{u}}\hat{r}}-\bar{\Lambda}^{\sf{\hat{u}}\sa} E^{\sa}_u\,,\\
0&=\partial_i \xi^u-\bar{\Lambda}^{\sf{\hat{u}}\sa}E^{\sa}_i\,,\\
0&=E^{\sf{\hat{r}}}_\alpha \partial_r \xi^\alpha-\bar{\Lambda}^{\sf{\hat{u}}\hat{r}}\,,\\
0&=E^{\sa}_\alpha \partial_r \xi^\alpha-\bar{\Lambda}^{\sf{\hat{u}}\sa}\,,
\end{align}
\end{subequations}
which are solved by
\begin{equation}
\partial_r \xi^u=0\,, \qquad \partial_r \xi^i=g^{ij}\partial_j \xi^u\,, \qquad \partial_r \xi^r=-\partial_u \xi^u+\left(E^{\sa}_i E^{\sa}_u-E^{\sf{\hat{r}}}_i\right) \partial_r \xi^i\,,
\end{equation}
together with
\begin{subequations}
\label{eq:Lambdas}
\begin{align}
\bar{\Lambda}^{\sf{\hat{u}}\sa}&=E^{\sa}_i \partial_r \xi^i\,,\\
\bar{\Lambda}^{\sf{\hat{u}}\hat{r}}&=\partial_r \xi^r+E^{\sf{\hat{r}}}_i \partial_r \xi^i\,.
\end{align}
\end{subequations}
The first set of equations implies
\begin{subequations}
\begin{align}
\xi^r&=-r \partial_u\xi^u+\alpha(u,x)+O(r^{-1})\,,\\
\xi^i&=Y^i(u,x)-r^{-1} \gamma^{ij} \partial_j \xi^u+O(r^{-2})\,,
\end{align}
\end{subequations}
and we deduce from \eqref{eq:Lambdas} that $\bar{\Lambda}^{\sf{\hat{u}}\sa}$ and $\bar{\Lambda}^{\sf{\hat{u}}\hat{r}}$ scale like $O(r^{-1})$ and $O(r^0)$ asymptotically, respectively. Looking at the transformation of the spin connection yields the additional restrictions
\begin{subequations}
\label{eq:residual.Omega}
\begin{align}
0&=\delta \Omega\indices{_r^{\sf{\hat{r}}\sa}}=\partial_r \Lambda^{\hat{r}\sa}-\xi^\mu \partial_r \Omega\indices{_\mu^{\sf{\hat{r}}\sa}}\,,\\
0&=\delta \Omega\indices{_r^{\sa \sf{b}}}=\partial_r \Lambda^{\sa \sf{b}}-\xi^\mu \partial_r \Omega\indices{_\mu^{\sa \sf{b}}}\,.
\end{align}
\end{subequations}
These are solved by LLT generators of the form
\begin{subequations}
\begin{align}
\Lambda^{\sf{\hat{r}}\sa}&=\lambda^{\sa}(u,x)+r^{-1}\left(\xi^u \omega_u^{(1)\sf{\hat{r}}\sa}+Y^i\omega_i^{(1)\sf{\hat{r}}\sa}\right)+O(r^{-2})\,,\\
\Lambda^{\sf{ab}}&=\lambda^{\sa \sf{b}}(u,x)+r^{-1}\left(Y^i\omega_i^{(1)\sf{ab}}\right)+O(r^{-2})\,,
\end{align}
\end{subequations}
where the leading expansion coefficients $\lambda^{\sa}$ and $\lambda^{\sa \sf{b}}$ are left unconstrained, while the subleading terms are fully determined by \eqref{eq:residual.Omega}. Having deduced the asymptotic scaling of all residual gauge parameters, one can check that preservation of the $E^{\sa}_u$ fall-off further requires
\begin{align}
\partial_u Y^i=0\,,
\end{align}
while all other fall-off conditions are trivially satisfied. 

Variations of the boundary metric and shear tensor are given by
\begin{subequations}
\begin{align}
\label{eq:gammavar}
\delta \gamma_{ij}&=2\left(-\partial_u\xi^u+\xi^u h_u\right) \gamma_{ij}+\lied_{Y}\gamma_{ij}\,,\\
\label{eq:Lied}
\delta C_{ij}&=\xi^u \partial_u  C_{ij}-\partial_u\xi^u C_{ij}+2\alpha \gamma_{ij}+\lied_{Y} C_{ij}-2 D_{(i} D_{j)} \xi^u\,,
\end{align}
\end{subequations}
where $\lied_{Y}$ is the intrinsic Lie derivative on the celestial sphere and $D$ is any metric compatible connection associated to $\gamma_{ij}$. Preservation of the shear tracelessness fully determines the function $\alpha$,
\begin{equation}
\alpha=\frac{1}{2}D^2\xi^u\,\,.
\end{equation}
The requirement that the determinant of $\bar{e}^{\sf{a}}_i$ be fixed implies
\begin{equation}
  \label{eq:detcondition}
 \delta (\sqrt{\gamma})=2\Theta^{-1}\,\delta \Theta\ \sqrt{\gamma}\,, 
\end{equation}
which, using \eqref{eq:gammavar}, yields
\begin{equation}
  \label{eq:frelation}
\omega\equiv \Theta^{-1}\delta \Theta=-\partial_u \xi^u+\xi^u h_u+ \frac{1}{2}D_iY^i \,.
\end{equation}
This equation can be interpreted in two different ways depending on the choice of independent gauge parameters, as emphasized in \cite{Barnich:2011ty}. One possibility would be to consider $\xi^u(u,x)$ an independent parameter, yielding the \emph{Newman--Unti group} as asymptotic symmetry group. Here, we choose instead to regard $\omega$ as one of the independent gauge parameters in order to single out boundary Weyl rescalings, which should be regarded as pure gauge. This alternative viewpoint yields the extended/generalized BMS group as asymptotic symmetry group. Equation \eqref{eq:frelation} then determines the $u$-dependence of $\xi^u$ in terms of $\omega$,
\begin{equation}
  \label{eq:82}
  \xi^u=\Theta\left(T(x)+\frac{1}{2}\int^u \dd u'\ \Theta^{-1}(D_iY^i-2\omega)\right)\,.
\end{equation}

In summary, the residual asymptotic symmetries are fully characterized by the arbitrary functions
\begin{equation}
  \label{eq:symmetryset}
\lbrace \omega(u,x),\ T(x),\ Y^i(x),\ \lambda^{\sa}(u,x),\ \lambda^{\sa\sf{b}}(u,x)\rbrace.
\end{equation}
Following earlier works \cite{Barnich:2010eb,Barnich:2011ty}, it is straightforward to compute the algebra of asymptotic Killing vectors $\xi[T,\omega,Y^i]$. Because of their explicit dependence on the metric field, one has to use the modified Lie bracket
\begin{equation}
[\xi_1,\xi_2]_M\equiv [\xi_1,\xi_2]-\delta^g_{\xi_1} \xi_2+\delta^g_{\xi_2} \xi_1\,,
\end{equation}
where $\delta^g_{\xi_1} \xi_2$ is the metric variation of $\xi_1$ induced from $\delta g_{\alpha\beta}=\lied_{\xi_1} g_{\alpha \beta}$. The resulting Lie algebra is
  \begin{equation}
  \label{eq:ASA} \Big[\xi[T_1,\omega_1,Y_1],\xi[T_2,\omega_2,Y_2]\Big]_M=\xi[T_{12},\omega_{12},Y_{12}]\,,
\end{equation}
with
\begin{subequations}
  \label{eq:89}
\begin{align} T_{12}&=Y_1^i\partial_iT_2+\frac{1}{2}T_1D_iY_2^i-(1\leftrightarrow 2)\,,\\
  Y_{12}^i&=Y_1^j \partial_j Y_2^i-(1\leftrightarrow 2)\,,\\
  \omega_{12}&=0\,.
\end{align}
\end{subequations}
By inspection of \eqref{eq:ASA}, one may recognize $\omega, T$ and $Y^i$ as the generators of boundary Weyl rescalings, supertranslations and superrotations -- diffeomorphisms of the celestial sphere --, respectively. Well-known algebras arise by considering additional phase space restrictions. For any fixed choice of $\Theta$, i.e.~for $\omega=0$, the above algebra reduces to the \textit{generalized BMS algebra} \cite{Campiglia:2014yka}. Requiring furthermore the metric $\gamma_{ij}$ to be conformal to the unit sphere metric in stereographic coordinates yields the \textit{extended BMS algebra} \cite{Barnich:2011mi}.\footnote{Vector fields generating Virasoro transformations have singularities on the celestial sphere, such that strictly speaking, they do not form a subset of Diff($\mathcal{S}^2$) generators.}

In the first order formulation of gravity which we are considering, we see the appearance of additional residual gauge parameters $\lambda^{\sa}$ and $\lambda^{\sa\sf{b}}$. As will be discussed in the next section, including these will allow us to identify a representation of the gauged $\mathfrak{so}(3,1)=\mathfrak{conf}(2)$ algebra at null infinity.

\section{Radiative vacua}
\label{sec:radiative-vacua}
In this section, we show that the subset of asymptotic fields $e^{\sa}, h, \omega^{\sf{ab}}, b^{\sa}$ introduced in \eqref{eq:framefall} and \eqref{eq:spinex} naturally falls into a representation of the gauged $\mathfrak{so}(3,1)$ algebra and may be consequently described in terms of a Chern--Simons theory. To show this, we first impose additional conditions on the phase space of section~\ref{sec:newman-unti-solution} that eliminate all radiative degrees of freedom. We organize the gravitational field equations into successive layers which may be solved iteratively near $\scri^+$. The fields $e^{\sa}, h, \omega^{\sf{ab}}, b^{\sa}$ are those appearing in the first layer of equations. In section \ref{sec:symmetrytrafos} we exhibit their SO(3,1) symmetry transformations before presenting a Chern--Simons theory which reproduces both symmetry transformations and field equations of the radiative vacuum sector in section \ref{sec:CStheory}. Finally, we complete the bulk analysis by solving the second layer of gravitational field equations in section \ref{sec:subleading}.

\subsection{Vacuum conditions}
\label{sec:vacconditions}
In order to exhibit a gauged $\mathfrak{so}(3,1)$ algebra, we impose the following three additional conditions on the phase space:
\begin{itemize}
    \item Vanishing of the leading order term in each component of $T^{\sf{\hat{r}}}$ and $T^{\sa}$,
    \item Vanishing of the leading order term of the scalar curvature $R$,
    \item Absence of \textit{gravitational radiation} at future null infinity.
\end{itemize}
The first two conditions are very mild as they are automatically satisfied by any gravitational solution satisfying the vanishing torsion constraint \eqref{eq:torsion} together with Einstein equations, provided that no matter source lies in a neighborhood of $\scri^+$. The last condition is much stronger and severely reduces  the number of physical situations that can be described. As far as an observer sitting at null infinity is concerned, this condition essentially restricts the states under consideration to gravitational vacua. This does not completely trivialize the discussion however, since gravitational vacua are known to be infinitely degenerate \cite{Strominger:2017zoo,Compere:2018aar,Compere:2016jwb,Compere:2018ylh}.

The torsion components of interest are given asymptotically by
\begin{subequations}
\begin{align}
T^{\sf{\hat{r}}}_{r\mu}&=\left(h_\mu-\omega\indices{_\mu^{\sf{\hat{u}}\hat{r}}}\right)+O(r^{-2})\,,\\
T^{\sf{\hat{r}}}_{\mu\nu}&=r\left(\partial_\mu h_\nu+b^{\sa}_\mu e^{\sa}_\nu- \mu \leftrightarrow \nu \right)\\
&+\left(\partial_\mu h^{(0)}_\nu+h_\mu h^{(0)}_\nu+b^{\sa}_\mu e^{(0)\sa}_\nu+\omega_\mu^{(1)\sf{\hat{r}}\sa} e^{\sa}_\nu-\mu \leftrightarrow \nu\right)+O(r^{-1})\,,\\
T^{\sa}_{r\mu}&=\left(e^{\sa}_\mu-\omega\indices{_\mu^{\sf{\hat{u}}\sa}}\right)+O(r^{-2})\,,\\
T^{\sa}_{\mu\nu}&=r\left(\partial_\mu e^{\sa}_\nu+\omega\indices{_\mu^{\sa \sf{b}}} e^{\sf{b}}_\nu+e^{\sa}_\mu h_\nu- \mu \leftrightarrow \nu \right)\\
&+\left(\partial_\mu e^{\sa(0)}_\nu+b^{\sa}_\mu \delta^u_\nu+e^{\sa}_\mu h^{(0)}_\nu+\omega\indices{_\mu^{\sf{ab}}}e^{(0)\sf{b}}_\nu-\mu \leftrightarrow \nu\right)+O(r^{-1})\,.
\end{align}
\end{subequations}
Setting the leading order terms to zero yields
\begin{equation}
\omega\indices{_\mu^{\sf{\hat{u}}\hat{r}}}=h_\mu, \qquad \omega\indices{_\mu^{\sf{\hat{u}}\sa}}=e^{\sa}_\mu\,,
\end{equation}
and
\begin{align}
\label{eq:dh}
0&=dh-e^{\sa} \wedge b_{\sa},\\
\label{eq:de}
0&=de^{\sa}+\omega\indices{^{\sa}_{\sf{b}}} \wedge e^{\sf{b}}+e^{\sa} \wedge h.
\end{align}
Here and in the following, the form notation is reserved for three-dimensional tensors living at null infinity.
As mentioned in the introduction, it is natural to interpret the fields $e^{\sa}$ as frame fields for the degenerate metric \eqref{eq:qzweibein} induced at $\scri^+$. The field $\omega^{\sa \sf{b}}$ appears as the spatial part of the associated spin connection. Equation \eqref{eq:de} then shows that it has non-vanishing torsion,
\begin{equation}
  \label{eq:torsion2d}
T^{\sa}\left(\omega\right)\equiv de^{\sa}+\omega\indices{^{\sa}_{\sf{b}}} \wedge e^{\sf{b}} =h \wedge e^{\sa},
\end{equation}
determined by $h$. However, defining a new connection $\hat{\omega}$ as in \eqref{eq:omegahat} shows that $h$ plays the role of the Weyl vector for the torsion-free Weyl connection $\hat{\omega}$. For more details on Weyl connections for Carrollian manifolds, see appendix~\ref{app:Weyl-connection}. The spatial components of equation \eqref{eq:de} is uniquely solved by
\begin{equation}
\label{eq:omega-solution}
\omega_i^{\sf{ab}}=-\omega_i \epsilon^{\sf{ab}}, \qquad \omega_i=\epsilon\indices{_{\sa}^{\sf{b}}} e^j_{\sf{b}} \left(\partial_i e^{\sa}_j-\partial_j e^{\sa}_i\right)+\epsilon\indices{_{\sa}^{\sf{b}}} e^{\sa}_i h_{\sf{b}}\,,
\end{equation}
while its $ui$ component holds identically thanks to the boundary condition \eqref{eq:hu}.

In order to investigate the second and third conditions, we first compute the Riemann tensor to the appropriate order,
\begin{subequations}
  \begin{align}
            \label{eq:domegau}
R_{ij}^{\sf{ab}}&=\partial_i \omega_j^{\sf{ab}}+b^{\sa}_i e^{\sf{b}}_j+e^{\sa}_i b^{\sf{b}}_j-(i\leftrightarrow j)+O(r^{-1}),
          \\
\label{eq:fluxequation}
    R_{\mu\nu}^{\sf{\hat{r}a}}&=\partial_\mu b^{\sf{a}}_\nu+h_\mu b^{\sf{a}}_\nu+\omega\indices{_\mu^{\sf{ab}}}b^{\sf{b}}_\nu-(\mu\leftrightarrow \nu)+O(r^{-1}),\\
      R_{ui}^{\sf{ab}}&,R_{\mu\nu}^{\sf{\hat{u}\hat{r}}} \sim O(r^{-1})\,,\\
    R_{r\mu}^{\sf{ab}}&, R_{r\mu}^{\sf{\hat{r}a}}, R_{\mu\nu}^{\sf{\hat{u}a}}\sim O(r^{-2})\,,\\
     R_{r\mu}^{\sf{\hat{u}\hat{r}}}&, R_{r\mu}^{\sf{\hat{u}a}}\sim O(r^{-3})\,.
\end{align}
\end{subequations}
It follows that the scalar curvature is given at leading order by
\begin{equation}
R=R^{\sA \sB}_{\alpha \beta} E_{\sA}^\alpha E_{\sB}^\beta=4r^{-2} e^i_{\sa} e^j_{\sf{b}} \left[\partial_i \omega_j^{\sf{ab}}+b^{\sa}_i e^{\sf{b}}_j+e^{\sa}_i b^{\sf{b}}_j-(i\leftrightarrow j) \right]+O(r^{-3}).
\end{equation}
Requiring that the leading term vanishes to order $r^{-2}$, we find
\begin{equation}
\label{eq:domega}
\dd \omega-\epsilon_{\sf{ab}}\ e^{\sa} \wedge b^{\sf{b}}=0\,.
\end{equation}
Note that the $ui$ component of this equation does not encode any additional information, since it follows directly from \eqref{eq:dh} and \eqref{eq:omega-solution}.
Assuming Einstein's equations, the scalar curvature of a geometry sourced by a stress-energy tensor with asymptotic falloff $r^{-s}$ vanishes like $r^{-(s+2)}$ \cite{Geroch1977}. Since finite energy configurations should have at least $s=2$, the above restriction on our phase space is very mild and compatible with physical solutions to Einstein's equations in asymptotically flat spacetimes.

The last requirement on our phase space is the absence of radiation. We will implement this condition by requiring that the the pull-back to $\scri^+$ of the \emph{magnetic part} of the (conformally rescaled) Weyl tensor $\tensor[^\star]{K}{^{\mu\nu}}$ vanishes \cite{Ashtekar:1981hw,Ashtekar:2014zfa}. It may be written
\begin{equation}
  \label{eq:starK}
\tensor[^\star]{K}{^{\mu\nu}}=-\lim_{r\rightarrow\infty}2r\,\Theta^2\epsilon^{\mu\kappa\lambda}C\indices{^{r\nu}_{\kappa\lambda}} \,,
\end{equation}
where we have defined the projected epsilon tensor at null infinity,
\begin{equation}
  \label{eq:90}
  \epsilon^{\mu\nu\rho}\equiv \lim_{r\rightarrow\infty}r^2\Theta^3\epsilon^{r\mu\nu\rho}\,.
\end{equation}
Although not apparent from these expressions, both quantities are  finite. In particular, the limit in \eqref{eq:starK} is well-defined since the Weyl tensor $C\indices{^{\alpha}_{\beta\gamma\delta}}$ vanishes to order $r^{-1}$ due to our boundary conditions. To leading order in $r$ we have
\begin{equation}
  \label{eq:88}
  E_{\sA}^r C\indices{^{\sf{A a}}_{\mu\nu}}=-\frac{1}{4}\Theta^{-2}\tensor[^\star]{K}{^{\nu \kappa}}\epsilon_{\lambda\mu\nu} e_\kappa^{\sa}\,.
\end{equation}
The leading order contribution on the left-hand side of this equation comes from the term $C\indices{^{\sf{\hat{r} a}}_{\mu\nu}}$.
Using Ricci decomposition for the left hand side of \eqref{eq:fluxequation} and checking that terms proportional to the Ricci tensor vanish to order $O(r^{-1})$, we can write
\begin{equation}
  \label{eq:newsequation}
  -\frac{1}{4}\Theta^{-2}\epsilon_{\mu\nu\rho}\tensor[^\star]{K}{^{\rho j}}\gamma_{j k}= 2 e^{\sa}_k\left(\partial_{[\mu} b^{\sf{a}}_{\nu]}+h_{[\mu} b^{\sf{a}}_{\nu}+\omega\indices{_{[\mu}^{\sf{ab}}}b^{\sf{b}}_{\nu]}\right)\,.
\end{equation}
Setting $\tensor[^\star]{K}{^{ui}}=\tensor[^\star]{K}{^{ij}}=0$ therefore yields
\begin{equation}
  \label{eq:db}
  \dd b^{\sf{a}}+h\wedge b^{\sf{a}}+\omega^{\sf{ab}}\wedge b^{\sf{b}}=0\,.
\end{equation}
In the Newman--Penrose formulation, the information contained in the magnetic part of the Weyl tensor is captured by the complex quantities $\textrm{Im}\, \psi_2^0\,,\psi_3^0\,,\psi_4^0$ \cite{Ashtekar:1981hw,Kozameh:1983aa}. The above condition corresponds to $\psi_3^0=\psi_4^0=0$ while leaving $\textrm{Im}\,\psi_2^0$ arbitrary.

\subsection{The Geroch tensor}
\label{sec:vaccondition}
In this section we expand on the role played by the field $b^{\sa}$. First, we observe that its $u$-component is completely determined by equation \eqref{eq:dh},
\begin{equation}
  \label{eq:busolution}
   b^{\sa}_{u}=e^{i}_{\sa}\left(\partial_u h_i-\partial_i h_u\right)\,.
\end{equation}
Considering its frame components
\begin{equation}
 \label{eq:beta-tensor}
\beta_{\sf{ab}}\equiv b^{\sa}_i e^i_{\sf{b}}\,,
\end{equation}
we find its antisymmetric and trace parts from \eqref{eq:dh} and \eqref{eq:domega},
\begin{equation}
  \label{eq:bequations}
 \beta_{[\sf{ab}]}=\frac{1}{2}e_{\sf{a}}^ie_{\sf{b}}^i\left(\partial_ih_j-\partial_j h_i\right)\,,\qquad \beta\equiv \beta_{\sf{ab}}\delta^{\sf{ab}}=\epsilon^{ij}\partial_i\omega_j=-\frac{1}{2}R(\omega)\,,
\end{equation}
while its traceless symmetric part remains so far undetermined.

Let us turn now to equation \eqref{eq:db}. In order to understand this equation better, we collect some results of the Geroch--Ashtekar covariant approach to asymptotically flat spacetimes \cite{Geroch1977,Ashtekar:1981hw}; see \cite{Ashtekar:2018lor} for a review. Note that these works employ the Bondi conformal frame and assume the standard metric compatible connection induced at the boundary.
Using Einstein's equations for the conformally compactified metric \eqref{eq:compactification} together with the Bianchi identities, one finds that the following equation holds at $\scri^+$,
\begin{equation}
  \label{eq:137} D_{[\mu}S_{\nu]\lambda}=\frac{1}{4}\epsilon_{\mu\nu\kappa}\tensor[^\star]{K}{^{\kappa\rho}}q_{\rho\lambda}\,.
\end{equation}
The pull-back $S_{\mu\nu}$ of the bulk Schouten tensor, defined in \eqref{eq:Schouten}, is a purely spatial tensor,
\begin{equation}
  \label{eq:51}
  n^{\mu}S_{\mu\nu}=0\,.
\end{equation}
Similarly, the operator $D_{\mu}$ is the Levi-Civitá derivative associated to $\tilde{g}_{\alpha\beta}$ pulled back to the boundary.
The tensor $S_{\mu\nu}$ plays a central role since it determines the presence of gravitational radiation as manifest in \eqref{eq:137}. However, it is not conformally invariant and thus depends on the choice of conformal factor $\Omega$ used to define the unphysical metric \eqref{eq:compactification}. The physical, conformally invariant content of $S_{\mu\nu}$ is contained in the \emph{News tensor}
\begin{equation}
  \label{eq:newsdefinition}
  N_{\mu\nu}\equiv S_{\mu\nu}-\rho_{\mu\nu}\,,
\end{equation}
where the symmetric tensor $\rho_{\mu\nu}$ obeys the three defining criteria
\begin{equation}
  \label{eq:rhocriteria}
n^{\mu}\rho_{\mu\nu}=0\,,\qquad D_{[\mu}\rho_{\nu]\lambda}=0\,,\qquad q^{\mu\nu}\rho_{\mu\nu}=R\left(q\right)\,.
\end{equation}
Here, $R\left(q\right)$ is the scalar curvature associated to the connection induced at the boundary. The Weyl transformation of the Geroch tensor $\rho_{\mu\nu}$ precisely cancels that of $S_{\mu\nu}$.
As shown in \cite{Geroch1977}, equations \eqref{eq:rhocriteria} define $\rho_{\mu\nu}$ uniquely when the topology of $\scri^+$ is $\mathbb{R}\times S^2$ and the vector fields generating asymptotic symmetries are regular everywhere; see also \cite{Campiglia:2020qvc}. Virasoro superrotation generators generically violate the latter requirement, and it is assumed that the phase space of asymptotically flat gravity should allow for more general topologies of $\scri^+$ in order to accommodate for solutions related by finite superrotations \cite{Strominger:2016wns,Adjei:2019tuj}. Note that from this point of view, the notion of News in this larger phase space is a priori ambiguous if no prescription for $\rho_{\mu\nu}$ is given. We consider that the News should vanish in the absence of gravitational radiation and that all remaining freedom is encoded in $\rho_{\mu\nu}$. Thus, the Geroch tensor can be identified with the \textit{vacuum Schouten tensor}. In the literature \cite{Compere:2018ylh,Compere:2016jwb}, this freedom has sometimes been absorbed into the News itself such that $\rho_{\mu\nu}|_{\text{here}}=\rho_{\mu\nu}-N_{\mu\nu}^{\text{vac}}|_{\text{there}}$.

Returning to our set-up, equation \eqref{eq:137} is clearly reminiscent of \eqref{eq:newsequation}, which can be rewritten as
\begin{equation}
  \label{eq:52}
   2\hat{D}_{[\mu}\beta_{i]k}=-\frac{1}{4\Theta^2}\epsilon_{\mu i\rho}\tensor[^\star]{K}{^{\rho j}}\gamma_{j k}\,,
\end{equation}
where $\hat{D}$ is the Weyl covariant derivative associated to $\hat{\omega}$. In the Bondi conformal frame $h=0$, an explicit computation yields
\begin{equation}
\label{eq:beta=Schouten}
\beta_{\mu\nu}=-\frac{1}{2} S_{\mu\nu}\,.
\end{equation}
This clarifies the role played by $b^{\sa}$. When restricting to radiative vacua satisfying $\tensor[^\star]{K}{^{\mu\nu}}=0$, it obeys the natural generalization to $h\neq 0$ of the relations \eqref{eq:rhocriteria} defining the Geroch tensor $\rho_{\mu\nu}$ (up to normalization). The undetermined purely spatial symmetric part of the field $b^{\sa}$ thus characterizes the various inequivalent radiative vacua. We will come back to the general determination of $b^{\sa}$ in section~\ref{sec:subleading} by solving Einstein's equations at subleading asymptotic order. In section~\ref{sec:solutionspace}, we will show that the solution space of radiative vacua is encoded in an effective action for superrotation reparametrization modes.

\subsection{Vacuum symmetry transformations}
\label{sec:symmetrytrafos}
We described radiative vacua in section~\ref{sec:vacconditions}  through equations \eqref{eq:dh}, \eqref{eq:de}, \eqref{eq:domega} and \eqref{eq:db} that hold near $\scri^+$ and which involve the asymptotic fields $e^{\sf{a}}, \omega^{\sf{ab}}, h, b^{\sf{a}}$ only. When the vacuum conditions are satisfied, their residual symmetry transformation may be written in a nicer form. Defining the gauge parameters
\begin{equation}
\eta \equiv \omega+Y^i h_i-\frac{1}{2} D_i Y^i\,, \qquad \pi^{\sa}\equiv Y^ie^{\sa}_i\,,
\end{equation}
 we straightforwardly obtain from \eqref{eq:gauge-transf-2} their residual gauge transformations in terms of the independent gauge parameters $\eta, \pi^{\sa}, \lambda^{\sa}, \lambda^{\sf{ab}}$,
\begin{subequations}
\label{onshell.sym}
\begin{align}
\label{onshell.sym:e}
\delta e^{\sa}&=\dd \pi^{\sa} - \pi^{\sa} h-\lambda\indices{^{\sa}_{\sf{b}}} e^{\sf{b}}+\eta e^{\sa}+\pi_{\sf{b}} \omega^{\sf{ab}},\\
\delta \omega^{\sf{ab}}&=\dd \lambda^{\sf{ab}}+\lambda\indices{^{\sf{b}}_{\sf{c}}}\omega\indices{_i^{\sf{ca}}}+\pi^{\sf{b}} b_i^{\sa}+\lambda^{\sf{b}} e^{\sa}_i - \left(\sa \leftrightarrow \sf{b} \right),\\
\delta h&=\dd \eta-\lambda_{\sa} e^{\sa}+\pi_{\sa} b^{\sa},\\
\delta b^{\sa}&=\dd \lambda^{\sa}+\lambda^{\sa} h -\lambda\indices{^{\sa}_{\sf{b}}} b^{\sf{b}}+\lambda_{\sf{b}} \omega\indices{^{\sf{ab}}}-\eta b^{\sa}\,.
\end{align}
\end{subequations}
The set of asymptotic fields $e^{\sf{a}}, \omega, h, b^{\sf{a}}$ seems to enjoy a particular status under these large gauge transformations, as they completely decouple from other subleading components in the asymptotic expansion of the bulk fields.
Together with the above mentioned field equations, these symmetry transformations show that a gauged $\mathfrak{so}(3,1)=\mathfrak{conf}(2)$ algebra governs the asymptotics of the vacuum phase space. As we will discuss in the next section, one can interpret this as a Chern--Simons theory based on the gauging of the symmetry algebra of the light cone at null infinity.

\subsection{Chern--Simons theory at null infinity}
\label{sec:CStheory}
In the seminal works \cite{Achucarro:1987vz,Witten:1988hc} it was shown that three-dimensional Einstein gravity in the first order formulation with zero, positive, or negative cosmological constant is classically equivalent to a Chern--Simons theory with gauge group $\textrm{ISO}(2,1)$, $\textrm{SO}(3,1)$, or $\textrm{SO}(2,2)$, respectively.
In general, a Chern--Simons theory is defined by a gauge group $G$, a non-degenerate invariant bilinear form $\langle\,,\rangle$ on the Lie algebra $\mathfrak{g}$ of $G$, and the action
\begin{equation}
  \label{eq:CSaction}
  S=\frac{k}{4\pi}\int\langle A\wedge \dd A+\frac{2}{3}A\wedge A\wedge A\rangle\,,
\end{equation}
where $A$ is a $\mathfrak{g}$-valued connection one-form. However, in order to relate the gauge field $A$ to the geometric fields of first-order gravity, namely a dreibein $e^{\sf{M}}$ and a spin connection $\omega^{\sf{MN}}$, an additional piece of data is needed. In particular, one has to choose a subalgebra $\mathfrak{h}$ of $\mathfrak{g}$. The components of $A$ associated to generators in this subalgebra are subsequently identified with the spin connection. For the cases mentioned above, one sets $\mathfrak{h}=\mathfrak{so}(2,1)$, i.e., the Lorentz group in three dimensions. The underlying structure providing the map between the Chern--Simons theory and its geometric interpretation in terms of vielbein and spin-connection is therefore the \emph{Klein pair} $(\mathfrak{g},\mathfrak{h})$ \cite{Matulich:2019cdo}, which defines a homogeneous space $G/H$.

Homogeneous spaces that obey certain physically motivated criteria, e.g., spatial isotropy, are called \emph{kinematic spacetimes}. These were classified in four dimensions in the seminal paper \cite{Bacry:1968zf} and more recently in any dimension in \cite{Figueroa-OFarrill:2018ilb}. The resulting spacetimes can be roughly divided in terms of their natural causal structure, i.e., Lorentzian, Galilean, or Carrollian. Chern--Simons theories based on the Klein pairs associated to three-dimensional Galilean and Carrollian spacetimes have been discussed in \cite{Papageorgiou:2009zc,Hartong:2016yrf,Bergshoeff:2017btm}.

Among the Carrollian spacetimes appearing in the classification of three-dimensional kinematical spacetimes of \cite{Figueroa-OFarrill:2018ilb} is the punctured future light-cone $\textrm{LC}_3$ seen as the homogeneous space
\begin{equation}
  \label{eq:93}
  \textrm{LC}_3=\textrm{SO}(3,1)/\textrm{ISO}(2)\,
\end{equation}
with topology $\mathbb{R}\times S^2$.
Note that the light-cone and three-dimensional de-Sitter are two very different homogeneous spacetimes for the same symmetry group, $\textrm{SO}(3,1)$, which shows the importance of the specification of the stabilizer algebra $\mathfrak{h}$.

Let $\{\Pt_{\sf{a}},\Ht,\Bt_{\sf{a}},\Jt\}$ be a basis for the algebra $\mathfrak{so}(3,1)$ with $\{\Bt_{\sf{a}},\Jt\}$ the generators for the subalgebra $\mathfrak{iso}(2,1)$. Using the conventions of \cite{Matulich:2019cdo}, the commutation relations for the light-cone algebra are then given by
\begin{subequations}
  \label{eq:LCalgebra}
\begin{align}
 [\Jt,\Bt^{\sf{a}}]&=\epsilon\indices{^{\sa}_{\sf{b}}}\Bt^{\sf{b}}\,, \qquad &&[\Jt,\Pt_{\sf{a}}]=\epsilon\indices{_{\sa}^{\sf{b}}}\Pt_{\sf{b}}\,,\\
[\Bt_{\sf{a}},\Pt_{\sf{b}}]&=\Ht\delta\indices{_{\sf{ab}}}-\Jt\epsilon\indices{_{\sf{ab}}}\,,\qquad &&[\Ht,\Pt_{\sf{a}}]=-\Pt_{\sf{a}}\,,\qquad &&[\Ht,\Bt^{\sf{a}}]=\Bt^{\sf{a}}\,.
\end{align}
\end{subequations}
We refer to \cite{Figueroa-OFarrill:2019sex,Duval:2014lpa,Duval:2014uva} for a thorough discussion of this algebra and the geometry of the light-cone regarded as a homogeneous space of the Lorentz group.

The commutation relations \eqref{eq:LCalgebra} may be interpreted in a different way as well. We can choose to take the quotient of $\textrm{SO}(3,1)$ by the subalgebra generated by $\{\Bt_{\sf{a}},\Jt,\Ht\}$ instead of that generated by $\{\Bt_{\sf{a}},\Jt\}$. The homogeneous space obtained in this way is the \emph{conformal two-sphere} $\textrm{CS}^2$
\begin{equation}
  \label{eq:134}
  \textrm{CS}^2=\textrm{SO}(3,1)/(\textrm{ISO}(2)\ltimes \mathbb{R})\,.
\end{equation}
In this context, $\Pt_{\sa}$, $\Ht$, $\Jt$, and $\Bt_{\sa}$ can be thought of as generators of translations, dilation, rotation, and special conformation transformations, respectively.
A Chern--Simons theory with this homogeneous space in mind was discussed in \cite{Gryb:2012qt}. Although the Chern--Simons theory and its equations are rather agnostic about the difference between \eqref{eq:93} and \eqref{eq:134}, we will see in the following that the latter choice appears to be more appropriate for interpreting the equations and symmetry transformations derived in the last two sections.

In addition to the gauge algebra, one needs an invariant bilinear form on the Lie algebra in order to define the theory. The most general invariant bilinear form for the algebra \eqref{eq:LCalgebra} is given by
\begin{align}
    \label{eq:invmetric}
  \langle \Jt,\Jt\rangle=-\langle \Ht,\Ht\rangle=\chi\,, \qquad \langle \Jt,\Ht\rangle=\mu\,, \qquad \langle \Pt_{\sf{a}},\Bt_{\sf{b}}\rangle=\mu\, \epsilon_{{\sf{ab}}}+\chi \delta_{\sf{ab}}\,.
\end{align}
While the choice of invariant bilinear form does not affect the classical theory, quantum theories based on different choices will be inequivalent in general. Since we do not have, at the level of our analysis, a preferred choice of invariant metric we will consider the most general one.
The Chern--Simons action is explicitly given by
\begin{align}
  \label{eq:CSLC}
  S&=\frac{k \mu}{2\pi}\int\left(\omega\wedge \dd h-\epsilon\indices{_{\sf{ab}}}b^{\sf{a}}\wedge \dd e^{\sf{b}}-\epsilon\indices{_{\sf{ab}}}e^{\sf{a}}\wedge b^{\sf{b}}\wedge h-e^{\sf{a}}\wedge b_{\sf{a}}\wedge \omega \right)\\
  \nonumber
     &+\frac{k \chi}{4\pi}\int\left(\omega\wedge \dd \omega-h\wedge \dd h-2b_{\sa}\wedge \dd e^{\sa}+2 h\wedge e^{\sa}\wedge b^{\sa}-2\epsilon_{\sf{ab}}\omega\wedge e^{\sa}\wedge b^{\sf{b}}\right)\,,
\end{align}
where the gauge field has been decomposed as
\begin{equation}
  \label{eq:CSconnection}
  A=h\Ht+e^{\sa}\Pt_{\sa}+b^{\sa}\Bt_{\sa}+\omega \Jt\,.
\end{equation}
Using the isomorphism $\textrm{SO}(3,1)\simeq \textrm{PSL}(2,\mathbb{C})=\textrm{SL}(2,\mathbb{C})/\mathbb{Z}_2$, the Chern--Simons action can be equivalently written as a sum of two complex conjugated actions \cite{Witten:1989ip}
\begin{equation}
  \label{eq:complexCS}
S=\frac{t}{8\pi}\int\langle A\wedge \dd A+\frac{2}{3}A\wedge A\wedge A\rangle_{\mathbb{C}}\,+\frac{\bar{t}}{8\pi}\int\langle \bar{A}\wedge \dd \bar{A}+\frac{2}{3}\bar{A}\wedge \bar{A}\wedge \bar{A}\rangle_{\mathbb{C}}\,
\end{equation}
with $t=k(\chi + i \mu)$, and where we take $A$ to be the $\mathfrak{sl}(2,\mathbb{C})$ connection
\begin{equation}
  \label{eq:Asl2C}
A=\left(e^{\sf{1}}+ie^{\sf{2}} \right)\Pt_{\sf{1}}+\left(h-i \omega \right)\Ht+\left(b^{\sf{1}}-ib^{\sf{2}} \right)\Bt_{\sf{1}}\,,
\end{equation}
with the usual invariant bilinear form on $\textrm{SL}(2,\mathbb{C})$
\begin{equation}
  \label{eq:bilinearC}
\langle \Bt_{\sf{1}},\Pt_{\sf{1}}\rangle_{\mathbb{C}}=\langle \Ht,\Ht\rangle_{\mathbb{C}}=1\,.
\end{equation}

Variation of the action yields the equations of motion
\begin{align}
  \label{eq:6}
 F(\Ht)=F(\Jt)=F(\Pt)=F(\Bt)=0\,,
\end{align}
where components of the field strength $F=\dd A+A\wedge A$ are given by
\begin{subequations}
  \label{eq:eomCS}
\begin{align}
  F(\Ht)&=\dd h-e^{\sf{a}}\wedge b_{\sf{a}}\,, \label{eq:RH}\\
  F(\Jt)&=\dd \omega-\epsilon\indices{_{\sf{ab}}}e^{\sf{a}}\wedge b^{\sf{b}}\,,\\
  F(\Pt)^{\sf{a}}&=\dd e^{\sf{a}}+e^{\sf{a}}\wedge h-\epsilon\indices{_{\sf{ab}}}\omega\wedge e^{\sf{b}}\,,\label{eq:RP}\\
  F(\Bt)_{\sf{a}}&=\dd b_{\sf{a}}-b_{\sf{a}}\wedge h-\epsilon\indices{_{\sf{ab}}}\omega \wedge b_b\,.
\end{align}
\end{subequations}
As anticipated, these field equations precisely agree with equations \eqref{eq:dh}, \eqref{eq:de}, \eqref{eq:domega} and \eqref{eq:db} that describe the phase space of four-dimensional asymptotically-flat radiative vacua. Similarly, gauge transformations of the Chern--Simons theory
\begin{equation}
  \label{eq:8}
  \delta_\Lambda A=\dd \Lambda+[A,\Lambda], \qquad \Lambda=\eta \Ht+\pi^{\sa}\Pt_{\sa}+\lambda^{\sa}\Bt_{\sa}+\lambda\Jt\,,
\end{equation}
precisely reproduce the symmetry transformations \eqref{onshell.sym} obtained in the last section. In order to find perfect agreement with the bulk gravitational theory, we must also assume that a $u$-coordinate labels the degenerate direction of the two Chern-Simons connection components $e^{\sa}$, and further impose \eqref{eq:hu} as boundary condition.
As a final comment let us mention that, taking the point of view that the theory is based on the gauging of the homogeneous space \eqref{eq:93}, the failure of the field equations \eqref{eq:eomCS} to determine $b^{\sa},\omega$ completely is well-known for Carrollian and Galilean theories \cite{Hartong:2015xda,Bergshoeff:2017btm}.

\subsection{Field equations at subleading order}
\label{sec:subleading}
In previous subsections we were mainly concerned with the first layer of equations found in an asymptotic expansion near $\scri^+$, that describe radiative vacua of asymptotically flat gravity. We now complete our bulk analysis by looking at the second layer of asymptotic equations. This allows us to discuss quantities that are commonly encountered in the literature.

Vanishing of the torsion to subleading order yields
\begin{equation}
\label{eq:dc}
\dd e^{(0)\sa}+\omega^{\sf{ab}} \wedge e^{(0)\sf{b}}-\dd u \wedge b^{\sa}-h^{(0)} \wedge e^{\sa}=0\,,
\end{equation}
which may be used to write the expression of $b^{\sa}$ in terms of frame field components,
\begin{subequations}
\label{eq:ba}
\begin{align}
b^{\sa}_u&=e^i_{\sa}\left(\partial_i h_u-\partial_u h_i\right),\\
b^{\sa}_{i}&=\frac{1}{2}\left(\partial_u C_{\sa \sf{b}}+h_u C_{\sa \sf{b}}-2h_u^{(0)} \delta_{\sf{ab}}-2D_{\sf{b}} h_{\sa}\right)e^{\sf{b}}_i\,.
\end{align}
\end{subequations}
Here, we introduced the two-dimensional covariant derivative $D$ associated to the boundary spin connection $\omega^{\sa \sf{b}}$,
\begin{equation}
D_{\sa} h_{\sf{b}}=e_{\sa}^i D_i h_{\sf{b}}=e_{\sa}^i \left(\partial_i h_{\sf{b}}+\omega\indices{_i_{\sf{b}}^{\sf{c}}} h_{\sf{c}}  \right).
\end{equation}
One can check that the solution \eqref{eq:ba} is compatible with \eqref{eq:dh}. Equation \eqref{eq:dc} is also used to find the spatial component of $h^{(0)}$,
\begin{equation}
 \label{eq:hi0}
h_i^{(0)}=-\frac{e^{\sf{b}}_i}{2} \left( D^{\sa} C_{\sf{ab}}+h^{\sa} C_{\sf{ab}} \right)=-\frac{1}{2} \left( D^k C_{ki}+h^k C_{ki} \right)\,,
\end{equation}
while its time component follows from \eqref{eq:domega},
\begin{equation}
h_u^{(0)}=-\frac{1}{2} \left(D_{\sa} h^{\sa}+\varepsilon^{ij} \partial_i \omega_j \right)=\frac{1}{4}\left(R(\omega)-2D_{\sa} h^{\sa}\right), \qquad \varepsilon_{ij}=e^{\sa}_i e^{\sf{b}}_j \epsilon_{\sa \sf{b}}\,.
\end{equation}
No other independent constraint on the second layer of fields $e^{(0)\sa}, h^{(0)}$ is derived from the bulk field equations evaluated near $\scri^+$. The fact that $\beta_{\mu\nu}$ defined in \eqref{eq:beta-tensor} is the Schouten tensor, is explicit from \eqref{eq:ba}. In the Bondi conformal frame $h=0$, we find
\begin{equation}
  \label{eq:53}
  \beta_{ij}=\frac{1}{2}\left(\partial_u C_{ij}-\frac{1}{2}R(\omega) q_{ij}\right)\,,
\end{equation}
where we recognize the contributions from the News tensor and the Geroch tensor $\rho_{\mu\nu}$.

As an interesting aside, we point out in appendix \ref{sec:extendedalgebra} that the field equations governing the first and second layers of fields $e^{\sa}, h, \omega, b^{\sa}$ and $e^{(0)\sa}, h^{(0)}$ may be recast as the vanishing of the field strength of a $\mathfrak{iso}(3,1)$-valued gauge connection. Similarly to what has been done in section~\ref{sec:symmetrytrafos}, we show that residual gauge transformations coincide with ISO(3,1) gauge transformations. However, it is well-known that these equations cannot follow from a three-dimensional Chern--Simons action due to a lack of nondegenerate ISO(3,1) bilinear form. Nevertheless, an action for the second layer of fields with the first layer considered a fixed background structure is found. This may prove useful if one is interested in constructing a boundary theory describing the shear tensor $C_{ij}$.

\section{Celestial Alekseev--Shatashvili theory}
\label{sec:reduct-bound-theory}
We have seen in the previous sections that the space of superrotation vacua can be described in terms of solutions of a Chern--Simons theory for the $\mathfrak{conf}(2)= \mathfrak{so}(3,1)$ algebra. Given the topological nature of Chern--Simons theories, it is natural to expect that it can be further reduced to a two-dimensional theory on the boundary of $\scri^+$, which we identify with the celestial sphere. The goal of this section is to do precisely this and derive an effective two-dimensional conformal field theory for superrotation vacua. We perform the reduction in section \ref{sec:reduction} and discuss the resulting theory and associated solution space in \ref{sec:solutionspace}.

\subsection{Hamiltonian reduction}
\label{sec:reduction}
The Hamiltonian reduction of Chern--Simons theory to a Wess--Zumino--Witten model on the boundary was pioneered in \cite{Elitzur:1989nr} and applied to the case of gravity on $\textrm{AdS}_3$ in \cite{Coussaert:1995zp}. More recently, the works \cite{Cotler:2018zff,Cotler:2019nbi,Merbis:2019wgk} revisited the reduction in the context of three-dimensional gravity on $\textrm{AdS}_3, \textrm{dS}_3$ and 3d flat space, respectively, from the point of view of the gravitational path integral. While we will not attempt to mirror the careful treatment of the path integral given in \cite{Cotler:2018zff,Cotler:2019nbi}, our approach closely follows these works to which we refer for further details.

In order to perform the reduction, it will prove most convenient to work with the form \eqref{eq:complexCS} of the Chern--Simons action. In what follows we will suppress the complex conjugated contribution and reinstate it only at the very end. The Hamiltonian form  of this Chern--Simons action is given by
\begin{equation}
  \label{eq:CSHam}
    S[A]=\frac{t}{8\pi}\int_{\mathcal \scri^+}\dd u \dd z \dd \bar{z}\ \langle A_u \partial_{\bar z} A_{z}- A_{z} \partial_{\bar z} A_{u}+2A_{\bar z} F_{z u}\rangle_{\mathbb{C}}\,,
\end{equation}
In the above we have introduced a coordinate system $(u,z,\bar z)$, where $z,\bar z$ are complex stereographic coordinates on the celestial sphere.
As we will see below, our choice of boundary conditions is such that no additional boundary term is needed in order to ensure a well-defined variational principle.

The field $A_{\bar{z}}$ appears as Lagrange multiplier for the constraint $F_{z u}=0$. Integrating out $A_{\bar z}$, the remaining fields must be flat connections that can be parametrized as
\begin{equation}
  \label{eq:111}
  A_u=G^{-1}\partial_u G\,, \qquad A_{z}=G^{-1}\partial_{z} G\,,
\end{equation}
with $G$ an arbitrary $\textrm{SL}(2,\mathbb{C})$ group element. Inserting these solutions of the constraint back into the action \eqref{eq:CSHam}, one finds after some integration by parts,
\begin{align}
  \label{eq:WZW}
  S&=\frac{t}{8\pi}\int_{\partial \scri^+}\dd z \dd \bar{z}\ \langle G^{-1}\partial_zGG^{-1}\partial_{\bar z} G\rangle_{\mathbb{C}}\\
  \nonumber
  &+\frac{t}{24\pi} \int_{\scri^+}\dd^3x\ \epsilon^{\mu\nu\lambda}\langle G^{-1}\partial_\mu G G^{-1}\partial_\nu G G^{-1}\partial_\lambda G\rangle_{\mathbb{C}}\,.
\end{align}
We have thus reproduced the well-known result that the Hamiltonian reduction of Chern--Simons theory yields a WZW theory on the boundary.

Before we continue, let us address two points regarding the reduction procedure and the topology of $\scri^+$. As already mentioned, when spacetimes related by finite superrotations are taken to be part of the phase space of asymptotically flat spacetimes, the celestial sphere acquires punctures. Consequently, one should allow for the possibility of holonomies in \eqref{eq:111}. We will disregard them in the present treatment and comment on their possible inclusion in the discussion.
Second, we note that the Chern-Simons theory could in principle be defined over any segment of $\scri^+$. Although any such segment has both a future and a past boundary, we restrict our attention to the past boundary only which we take to be $\scri^+_-$ for simplicity. As will be shown in a moment, the Hamiltonian reduction yields time-independent field configurations such that the actual choice of boundary is immaterial. Finally, one should in principle treat both future and past boundaries on the same footing in performing the reduction. The presence of holonomies presumably leads to couplings between both boundaries along the lines of \cite{Cotler:2018zff,Henneaux:2019sjx}. We will leave this interesting question for future investigation.

In order to proceed, we choose a parametrization for the group element $G$. Using the Gauss decomposition
 \begin{equation}
   \label{eq:Gpar}
   G=e^{\Pi \Pt_{\sf{1}}}e^{\Lambda \Ht}e^{B \Bt_{\sf{1}}}\,,
 \end{equation}
 with $\Pi,\Lambda,B$ being complex functions of the coordinates $(u,z,\bar{z})$, we find
\begin{equation}
  \label{eq:Gcurrent}
  G^{-1}\dd G=\left(e^{\Lambda}\dd \Pi\right)\Pt_{\sf{1}}+\left(\dd \Lambda-e^{\Lambda}B \dd \Pi\right)\Ht+\left(\dd B+B \dd \Lambda-\frac{1}{2}e^{\Lambda}B^2\dd \Pi\right)\Bt_{\sf{1}}\,,
\end{equation}
and
\begin{equation}
  \label{eq:WZWcontribution}
  \langle (G^{-1}\dd G)^3\rangle_{\mathbb{C}}=\dd\left(e^{\Lambda}\dd \Pi\wedge \dd B\right)\,.
\end{equation}
Other parametrizations than \eqref{eq:Gpar} could a priori be considered, and should in principle correctly account for the presence or absence of holonomies; see for example \cite{Cotler:2018zff,Cotler:2019nbi,Henneaux:2019sjx} in the context of (A)dS$_3$ gravity. As mentioned above, we leave a precise study of holonomies to future work, and simply check below that the parametrization \eqref{eq:Gpar} is such that the group element corresponding to Minkowski space is single-valued on the celestial sphere.

At the moment, it appears that we obtain a $\textrm{SL}(2,\mathbb{C})$ WZW model on the boundary. We push the reduction further by restricting to the conformal gauge 
\begin{equation}
  \label{eq:conformal-gauge}
  e^{\sa}e^{\sa}=\Theta^2 \gamma_{z\bar{z}} \dd z \dd \bar z\,, \qquad \gamma_{z\bar{z}}=\frac{4}{(1+z \bar{z})^2}\,.
\end{equation}
This condition explicitly breaks Diff($\mathcal{S}^2$) superrotation symmetry. In order to simplify formulas, in the following we define $\gamma\equiv \sqrt{\gamma_{z\bar{z}}}$. The above gauge condition is then implemented by
\begin{equation}
\label{eq:WZWcon1}
e^{\sf{1}}+i e^{\sf{2}}= \Theta \gamma \dd z, \qquad \omega_z=-i \partial_z \ln \gamma \,.
\end{equation}
In addition, we require the Weyl vector to be pure gauge,
\begin{equation}
\label{eq:WZWcon2}
h_\mu=\partial_\mu \ln \Theta.
\end{equation}
The $u$ component of this equation is the boundary condition \eqref{eq:hu}, while the $z, \bar{z}$ components are gauge conditions which we impose. Note that we are not allowed to assume a $u$-independent $\Theta$ in accordance with the comment regarding Bondi conformal frames below \eqref{eq:81}. With these restrictions, it is now straightforward to check that no surface term is needed in \eqref{eq:CSHam}.

The conditions \eqref{eq:WZWcon1} and \eqref{eq:WZWcon2} impose restrictions on the group element $G$ or, equivalently, current constraints in the WZW model. We find
\begin{align}
  \label{eq:130}
  \Lambda=\ln \left(\frac{ \Theta\gamma}{\partial_ z \Pi}\right),\qquad \partial_u \Pi=0, \qquad B=\left(\Theta\gamma\right)^{-1}\left(2\, \partial_z\ln \gamma-\frac{\partial^2_{z}\Pi}{\partial_ z\Pi}\right)\,,
\end{align}
such that all fields are written in terms of $\Pi=\pi^{\sf{1}}+i\pi^{\sf{2}}$, a complex form of the superrotation symmetry generators $\pi^{\sa}=e^{\sa}_i Y^i$.

We need to check that the parametrization \eqref{eq:Gpar} properly describes Minkowski space, and in particular that the associated group element is single-valued on the celestial sphere. This corresponds to $\Pi(z)=z$, such that together with the constraints \eqref{eq:130} and the stereographic projection map $z=\tan(\theta/2)e^{i \phi}$, one finds the corresponding group element
\begin{equation}
G=
\begin{pmatrix}\sqrt{1+\cos\theta}&-e^{i \phi}\sin\frac{\theta}{2}\\e^{i \phi}\sin\frac{\theta}{2}&\frac{1}{2}\sqrt{1+\cos\theta}
\end{pmatrix}.
\end{equation}
This coordinate system makes it manifest that the above group element is single-valued on the sphere as required. As already mentioned above, for generic field configurations $\Pi(z)$, which can be interpreted as introducing punctures on the celestial sphere, the issue of holonomies becomes more subtle and requires a better understanding of the phase space of superrotation vacua.

Returning to the Hamiltonian reduction, plugging the above constraints \eqref{eq:130} into the action \eqref{eq:WZW}, and after integrating by parts, dropping field-independent terms involving $\Theta$ and $\gamma$ only, and reinstating the complex conjugate contribution to \eqref{eq:CSHam}, one finds
\begin{equation}
  \label{eq:final}
  S\left[\Pi\right]=\frac{t}{16\pi}\int_{\mathcal{S}^2}\dd z \dd \bar{z}\
  \frac{\partial_ z \partial_{\bar{z}}\Pi\ \partial_z^2\Pi}{(\partial_z \Pi)^2}+\frac{\bar{t}}{16\pi}\int_{\mathcal{S}^2}\dd z \dd \bar{z}\
  \frac{\partial_{\bar{z}}\partial_z\bar{\Pi}\ \partial_{\bar{z}}^2\bar{\Pi}}{(\partial_{\bar{z}} \bar{\Pi})^2}\,.
\end{equation}
This effective action for the superrotation mode $\Pi$ is the main result of the present work. It is recognized as a complex version of the Alekseev-Shatashvili geometric action on a coadjoint orbit of the Virasoro group \cite{Alekseev:1988ce}.

The action \eqref{eq:final} is invariant under two different sets of symmetries. Transformations of the form
\begin{equation}
  \label{eq:symAS}
  \delta_\epsilon \Pi=\epsilon(z) \partial_z \Pi\,,
\end{equation}
are genuine symmetries of the action and yield an infinite number of conserved charges
\begin{equation}
  \label{eq:chargeAS}
  Q[\epsilon]=\int \dd z\,\epsilon(z) T(z)\,,
\end{equation}
where $T$ is the Schwarzian derivative of the superrotation mode,
\begin{equation}
  \label{eq:Schwarzian}
T=-\frac{1}{2} \lbrace \Pi;z \rbrace\,, \qquad \lbrace \Pi;z \rbrace=\frac{\partial_z^3 \Pi}{\partial_z \Pi}-\frac{3}{2} \left(\frac{\partial_z^2 \Pi}{\partial_z \Pi}\right)^2\,.
\end{equation}
The other symmetry transformation is pure gauge and acts as
\begin{equation}
  \label{eq:gaugeAS}
  \Pi \mapsto \frac{a \Pi+b}{c \Pi +d}\,, \qquad ad-bc=1\,, \qquad a,d,b,c \in \mathbb{C}\,,
\end{equation}
which is recognized as a finite $\textrm{PSL}(2,\mathbb{C})$ transformation. It can be traced back to the invariance of \eqref{eq:111} under transformations $G\mapsto g G$, with $g$ a general ($\bar{z}$-dependent) $\textrm{PSL}(2,\mathbb{C})$ element. This redundancy in the reduction reappears as the gauge symmetry \eqref{eq:gaugeAS}.

The action \eqref{eq:final} (with $t$ imaginary) has been recently related to the path integral quantization of gravity on Euclidean $\textrm{AdS}_3$ and Lorentzian $\textrm{dS}_3$ \cite{Cotler:2018zff,Cotler:2019nbi}. Given the fact that these theories are also based on a $\textrm{PSL}(2,\mathbb{C})$ Chern--Simons theory, its reappearance in this context is quite natural. We will not go into further details regarding the interesting properties of the Alekseev-Shatashvili action and its relation to coadjoint orbits of the Virasoro group, and refer the reader to the thorough discussions in \cite{Alekseev:1988ce,Barnich:2017jgw,Cotler:2018zff}.

\subsection{Phase space of superrotation vacua}
 \label{sec:solutionspace}
 We now examine solutions of the Alekseev-Shatashvili action \eqref{eq:final}, which turn out to describe the Virasoro superrotation vacua described in \cite{Compere:2016jwb}.

The equation of motion of the Alekseev-Shatashvili action simply is
\begin{equation}
  \label{eq:eomAS}
\partial_{\bar{z}} T=0\,,
\end{equation}
and is equivalent to the statement that the charges \eqref{eq:chargeAS} are conserved.
It is solved by
\begin{equation}
  \label{eq:solAS}
\partial_{\bar{z}}\Pi=0\,,
\end{equation}
i.e., $\Pi$ ($\bar{\Pi}$) is a holomorphic (anti-holomorphic) function. Had we solved the Chern--Simons theory directly without going through the Hamiltonian reduction, we would have imposed $A_{\bar z}=G^{-1}\partial_{\bar z}G$ as solution to the constraints in addition to \eqref{eq:111}. Equation \eqref{eq:solAS} would have followed from the boundary conditions \eqref{eq:WZWcon1}, thus showing consistency of the reduction procedure.

Together with the results of the previous section we can now use the solutions of the Alekseev-Shatashvili theory to characterize the superrotation vacua associated to the metric \eqref{eq:conformal-gauge}.
First notice that reality of the 1-forms $b^{\sa}$ in stereographic coordinates implies $(b^{\sa}_{z})^*=b^{\sa}_{\bar{z}}$. From the definition \eqref{eq:beta-tensor},
we find
\begin{subequations}
\begin{align}
\beta_{zz}&=\frac{\Theta \gamma}{2}\left(b^{\sf{1}}_z-i b^{\sf{2}}_z\right), \qquad \beta_{\bar{z}\bar{z}}=(\beta_{zz})^*\,,\\
\beta_{z\bar{z}}&=\frac{\Theta \gamma}{2}\left(b^{\sf{1}}_{\bar{z}}-i b^{\sf{2}}_{\bar{z}}\right), \qquad \beta_{\bar{z}z}=(\beta_{z\bar{z}})^*\,.
\end{align}
\end{subequations}
Finally, using \eqref{eq:111} together with the constraints \eqref{eq:130} and the solution \eqref{eq:solAS}, we find
\begin{equation}
  \label{eq:vacuabeta}
\beta_{zz}=T+ \frac{1}{2} \partial_z^2 \ln \gamma-\frac{1}{4}\left(\partial_z \ln \gamma \right)^2\,,
\end{equation}
and
\begin{equation}
\beta_{\bar{z}z}=\partial_z \partial_{\bar{z}} \ln \gamma\,, \qquad \beta_{z\bar{z}}=\beta_{\bar{z}z}\,.
\end{equation}
The trace of the Geroch tensor in stereographic coordinates is given by
\begin{equation}
\label{eq:trace-beta}
\beta=4 \left( \Theta \gamma\right)^{-2} \beta_{z\bar{z}}=-\frac{1}{2} R\left(\omega\right)\,,
\end{equation}
consistent with the general solution \eqref{eq:bequations}. The set of vacuum configurations $\rho_{\mu\nu}=-2\beta_{\mu\nu}|_{\text{vac}}$ is parametrized by the Schwarzian $T$ of the Virasoro superrotation mode $\Pi(z)$, which therefore labels the space of superrotation vacua.

It is instructive to reconstruct the bulk metric corresponding to these superrotation vacua. For convenience we first perform a Weyl rescaling $\Theta \to 1$, and subsequently use \eqref{eq:53} to solve for $C_{ij}$ in terms of $\beta_{ij}$ given by \eqref{eq:vacuabeta}.
Plugging this back into \eqref{eq:NUmetric}, we obtain
\begin{equation}
  \label{eq:superrotationvacua}
\dd s^2=-\dd u^2-2 \dd u \dd r+r^2 \gamma_{z \bar{z}}\dd z \dd \bar{z}+ r (C_{zz} \dd z^2+C_{\bar{z}\bar{z}} \dd \bar{z}^2)+...\,,
\end{equation}
with
\begin{equation}
  \label{eq:vacuumC}
C_{zz}=2 u T+ C^{T}_{zz},\qquad \partial_u C^{T}_{zz}=0\,.
\end{equation}
The subleading orders are completely fixed in terms of the functions given above.
The function $C^{T}_{zz}$ parametrizes the space of \emph{supertranslation} vacua which we cannot access from the perspective of the Alekseev-Shatashvili theory. We thus reproduce the metric of Virasoro superrotation vacua as previously constructed in \cite{Compere:2016jwb}; see also \cite{Adjei:2019tuj}. A similar result was derived in \cite{Barnich:2016lyg}, where it was shown that the transformation of the shear tensor $C_{zz}$ under Virasoro superrotations involves the Schwarzian derivative of the symmetry parameter $\Pi(z)$. In fact, the results of the present work can be employed to reproduce the finite Weyl and superrotation transformations of \cite{Barnich:2016lyg} using the finite transformations of CS fields under the gauge group $\mathfrak{so}(3,1)$.

\section{Discussion}

We have derived an effective action for the superrotation reparametrization mode $\Pi(z,\bar{z})$ in the form of a complex Alekseev--Shatashvili action. Classical solutions of this theory are holomorphic configurations $\Pi(z)$ and precisely correspond to the \textit{Virasoro superrotation vacua} described in \cite{Compere:2016jwb}, which spontaneously break the Virasoro symmetry group down to its global conformal subgroup. Holomorphic configurations $\Pi(z)$ are flat directions of the Alekseev--Shatashvili theory as they have zero action. On the other hand, generic configurations $\Pi(z,\bar{z})$ have non-zero action and correspond to generators of Diff($\mathcal{S}^2$) superrotation symmetries that are explicitly broken by a choice of conformal gauge at $\scri^+_-$.

As pointed out above, in the derivation of our result we did not include a holonomy term in \eqref{eq:111}. Nevertheless, consider for instance the configuration $\Pi(z)=z^{\alpha}$ with $\alpha$ not an integer. At the level of our analysis, it is a solution of the equation of motion \eqref{eq:solAS}. However, it is straightforward to check that the holonomy of the associated group element is non-trivial and the group element is not single-valued. This is not surprising as these transformations are known to generate defects in the bulk -- interpreted as cosmic strings in \cite{Strominger:2016wns} -- or at the boundary \cite{Adjei:2019tuj}, depending on the point of view taken. It is thus clear that our assumption, while useful as a first step, should be revisited in order to address these important issues regarding the topology of null infinity.

In the present approach, we introduced the Chern--Simons theory, which eventually led to our main result \eqref{eq:final}, as an effective description for the symmetries and equations of motion of Einstein gravity near $\scri^+$. In particular, we did not derive the Chern--Simons theory by reducing the Einstein--Hilbert action, restricted to our phase space of vacuum solutions, to the boundary. Therefore, we have no information about the coupling constant $t$ in the final theory that is related to the central charge of the corresponding conformal field theory (CFT). This should be contrasted with similar work \cite{Cotler:2018zff,Cotler:2019nbi,Merbis:2019wgk,Carlip:2016lnw} in lower dimensions where boundary actions were directly derived from the bulk action, or work in four dimensions which showed that the action for gravity reduces to a Chern--Simons theory on the horizon of an isolated black hole \cite{Ashtekar:1999wa,Basu:2010hv}. However, the fact that our result is the natural generalization of \cite{Carlip:2016lnw} to four dimensions suggests that it should be possible to arrive at our result by reducing the Einstein--Hilbert action.  The central charge would then likely be inversely proportional to Newton's constant. We leave open this important and interesting problem. It would also be interesting to see whether the central charge is related to the field-dependent central terms appearing in \cite{Barnich:2010eb}.

The Alekseev-Shatashvili action \eqref{eq:final} is closely related to the 2d induced gravity theory of Polyakov \cite{Polyakov:1987zb}
which reduces to the former upon a particular choice of gauge. It is an intriguing possibility that the Polyakov action would turn out to be the effective action of superrotations if we were not to impose the gauge-fixing condition \eqref{eq:conformal-gauge} when performing the Hamiltonian reduction of the Chern--Simons theory, thereby keeping Diff($\mathcal{S}^2$) symmetry manifest. A likely related matter is the appearance of a Liouville action (partly) describing the \textit{smooth} Diff$(\mathcal{S}^2)$ superrotation vacua \cite{Compere:2018ylh}. The Liouville field $\Phi$ may be identified with the celestial conformal factor through $\Theta=e^{-\Phi/2}$. Indeed, equation \eqref{eq:trace-beta} may be shown to coincide with the Liouville equation $D^2 \Phi=R$.\footnote{After this paper was written, the various relations between the Alekseev--Shatashvili, Polyakov and Liouville actions, together with their role as generating functionals for CFT stress tensor correlators, have been clarified in \cite{Nguyen:2021dpa}.}

Finally, it would be of great interest to connect the celestial Alekseev-Shatashvili theory derived here with recent work on the soft sector of celestial CFT amplitudes \cite{Donnay:2018neh,Donnay:2020guq,Puhm:2019zbl,Himwich:2020rro}. Another promising research avenue would be to exploit the well-known connection between $\textrm{SL}(2,\mathbb{C})$ Chern--Simons theories and CFT \cite{Verlinde:1989ua} in order to shed further light on gravity. We leave this to future endeavors.

\section*{Acknowledgments}
We thank Stefan Prohazka, Max Riegler, Romain Ruzziconi and Andy Strominger for interesting discussions on related topics. JS thanks Roberto Emparan for discussions and Jordan Cotler, Kristan Jensen, Stefan Prohazka, and Max Riegler for collaboration on related work. The work of KN is supported by a Fellowship of the Belgian American Educational Foundation and by a grant from the John Templeton Foundation.  The work of JS is supported by the Erwin-Schr\"odinger fellowship J-4135 of the Austrian Science Fund (FWF), and by the ERC Advanced Grant GravBHs-692951 during the early stages of this work.

\appendix
\section{Weyl connections on Carrollian manifolds}
\label{app:Weyl-connection}
In this appendix we develop the concept of Weyl connections on Carrollian manifolds in more details. Connections on Carrollian manifolds have been treated in, e.g., \cite{Duval:2014lpa,Bekaert:2015xua}. Weyl connections on Carrollian manifolds have been discussed from a different perspective in \cite{Ciambelli:2018xat,Ciambelli:2018wre,Ciambelli:2018ojf}.

Only in this subsection we change our conventions such, that indices $\mu,\nu,...$ stand for $D$-dimensional coordinate indices, $A,B,C,...$ denote $D$-dimensional frame indices and $a,b,c,..$ denote frame indices associated to the non-degenerate $(D-1)$-dimensional Riemannian submanifold.

A Carrollian geometry is characterized by a degenerate metric $q_{\mu \nu}$ whose kernel is spanned by the vector $n^\mu$,
\begin{equation}
n^\mu q_{\mu\nu}=0.
\end{equation}

A Weyl covariant derivative $D$ is defined from the requirement that angles between vectors be preserved under parallel transport along any curve with tangent vector $t$,
\begin{equation}
\label{angle.preservation}
t^\mu D_\mu \left(\frac{v.w}{|v||w|}\right)=0\,, \qquad \forall\ v,w \quad \text{s.t.} \quad t^\mu D_\mu v^\nu=t^\mu D_\mu w^\nu=0\,.
\end{equation}
We find
\begin{align}
t^\rho D_\rho \left(\frac{v.w}{|v||w|}\right)=t^\rho D_\rho q_{\mu\nu} \left(\frac{v^\mu w^\nu}{|v||w|}-\frac{1}{2} \frac{v.w}{|v||w|}\left(\frac{v^\mu v^\nu}{v^2}+\frac{w^\mu w^\nu}{w^2}\right)\right)\,,
\end{align}
such that condition \eqref{angle.preservation} is equivalent to
\begin{equation}
\label{Weyl.metricity}
D_\rho q_{\mu\nu}=2h_\rho q_{\mu\nu}\,,
\end{equation}
where the \textit{Weyl vector} $h$ is a priori arbitrary. In particular, this connection is not metric compatible. We further impose invariance of the covariant derivative $D$ and covariance of the condition \eqref{Weyl.metricity} under Weyl rescalings. Hence, the rescaled metric
\begin{equation}
\tilde{q}=e^{2\eta} q\,,
\end{equation}
should satisfy
\begin{equation}
\tilde{D}_\rho \tilde{q}_{\mu\nu}=2\tilde{h}_\rho \tilde{q}_{\mu\nu}\,,
\end{equation}
or equivalently,
\begin{equation}
\tilde{D}_\rho q_{\mu\nu}=2\left(\tilde{h}_\rho- \partial_\rho \eta\right)q_{\mu\nu}\,.
\end{equation}
Invariance of the covariant derivative yields
\begin{align}
0=D_\rho q_{\mu\nu} -\tilde{D}_\rho q_{\mu\nu}=2\left(h_\rho -\tilde{h}_\rho+ \partial_\rho \eta \right)q_{\mu\nu}\,,
\end{align}
which determines the transformation of the Weyl vector under Weyl rescalings,
\begin{equation}
\tilde{h}_\rho=h_\rho +\partial_\rho \eta\,.
\end{equation}
The above considerations actually apply to manifolds equipped with both degenerate or non-degenerate metrics. In the case of a degenerate metric, it is natural to impose that parallel transport also preserves the metric degeneracy, 
\begin{equation}
0=t^\rho D_\rho \left(n^\mu q_{\mu\nu}\right)=t^\rho \left(q_{\mu\nu} D_\rho n^\mu+n^\mu D_\rho q_{\mu\nu}\right)=t^\rho q_{\mu\nu} D_\rho n^\mu\,.
\end{equation}
The solution to this equation is of the form
\begin{equation}
\label{Weyl.metricity.2}
D_\mu n^\nu=w_\mu n^\nu\,,
\end{equation}
where the vector $w$ is also not specified a priori. The behavior of $w$ under Weyl rescaling may be inferred from the Weyl transformation of the normal vector $n$. In the covariant description of null infinity \cite{Geroch1977,Ashtekar:2014zsa}, one usually considers the following transformation properties,
\begin{equation}
\tilde{n}^\mu=e^{-\eta} n^\mu\,, \qquad \tilde{w}_\mu=w_\mu-\partial_\mu \eta\,.
\end{equation}
Finally, one can check that
\begin{equation}
\lied_n q_{\mu\nu}=n^\lambda D_\lambda q_{\mu\nu}+q_{\nu\lambda} D_\mu n^\lambda+q_{\mu\lambda} D_\nu n^\lambda=2 (n.h) q_{\mu\nu}\,.
\end{equation}
Hence, a non-zero normal component of the Weyl vector is associated to volume changes along the degenerate direction.

\paragraph{Affine Weyl connection.}
Let's turn to the affine connection $\Gamma$ associated to a Weyl covariant derivative. For generality, we work with a generic (non-holonomic) vector basis  $\underline{e}_{A}=\left(n,\underline{e}_{a}\right)$, where the basis vector $n\equiv e_n$ is chosen to lie along the degenerate direction. The affine connection is defined by
\begin{equation}
  \label{eq:Riccirot}
D_{A} \underline{e}_{B}\equiv \Gamma_{AB}^{C} \,\underline{e}_{C}\,.
\end{equation}
In case that the basis vectors are associated to a coordinate system (holonomic), $\underline{e}_A=\delta_A^\mu \partial_\mu$, the connection coefficients coincide with the usual Christoffel symbols. We will restrict our attention to torsionless connections, in which case we have
\begin{equation}
\Gamma_{AB}^{C}-\Gamma_{BA}^{C}=\left[\underline{e}_{A},\underline{e}_{B}\right]\cdot \underline{e}^{C}\equiv C\indices{_{AB}^{C}}\,.
\end{equation}
In this basis, the Weyl metricity condition \eqref{Weyl.metricity} yields
\begin{align}
\underline{e}_{C}\left(q_{AB}\right)-\Gamma_{CA}^{D} q_{BD}-\Gamma_{CB}^{D} q_{AD}=2 h_{C} q_{AB}\,,
\end{align}
such that
\begin{align}
\label{Gamma.nonholonomic}
  \Gamma_{AB}^{D} q_{CD}&=\frac{1}{2}\left(\underline{e}_{A}(q_{BC})+\underline{e}_{B}(q_{AC})-\underline{e}_{C}(q_{AB})\right)\\
\nonumber
&-\frac{1}{2}\left(C\indices{_{AC}^{D}}q_{BD}+C\indices{_{BC}^{D}} q_{AD}-C\indices{_{AB}^{D}} q_{CD}\right)-\left(h_{A} q_{BC}+h_{B} q_{AC}-h_{C} q_{AB} \right)\,.
\end{align}
The other condition \eqref{Weyl.metricity.2} yields
\begin{equation}
\Gamma_{A n}^{C}=w_{A}\ \delta^{C}_{n}\,,
\end{equation}
or equivalently,
\begin{equation}
  \label{connection.condition.2}
\Gamma_{An}^{n}=w_{A}, \qquad \Gamma_{An}^{a}=0\,.
\end{equation}
Together with \eqref{Gamma.nonholonomic}, this last condition implies
\begin{equation}
h_n\ q_{ab}=\frac{1}{2}\left(n(q_{ab})+C\indices{_{a n}^{c}} q_{cb}+C\indices{_{bn}^{\sc}} q_{ac}\right)\,.
\end{equation}
In a coordinate basis where $n=\partial_u,\ \underline{e}_{a}=\delta_{a}^i \partial_i$, this equation simplifies to
\begin{equation}
\partial_u q_{ij}=2 h_u q_{ij}\,.
\end{equation}
We note here that, in contrast to Levi-Civitá connections for non-degenerate metrics, equation \eqref{Gamma.nonholonomic} does not determine the connection coefficients completely.

\paragraph{Spin Weyl connection.}
We specialize to the particular case where the basis vectors are also frame fields on the degenerate manifold,
\begin{equation}
\label{Weyl.frame}
q_{\mu\nu}=\delta_{ab} e^{a}_\mu e^{b}_\nu\,.
\end{equation}
In this case, the Weyl spin connection $\hat{\omega}$ is identified with
\begin{equation}
\hat{\omega}\indices{_{A}^{C}_{B}}\equiv \Gamma_{AB}^{C}\,.
\end{equation}
From \eqref{Weyl.frame},
we find
\begin{align}
D_\rho q_{\mu\nu}=-\delta_{ab}\hat{\omega}\indices{_\rho^{b}_{C}}\left( e^{C}_\mu e^{a}_\nu+e^{a}_\mu e^{C}_\nu \right)\,,
\end{align}
such that \eqref{Weyl.metricity} implies
\begin{equation}
  \label{spin.connection.1}
  \hat{\omega}_{\mu (ab)}=-h_\mu \delta_{ab}\,, \qquad \hat{\omega}\indices{_\mu^{a}_{n}}=0\,.
\end{equation}
Hence, its projection onto the non-degenerate submanifold differs from the Levi-Civit\'a spin connection $\omega^{ab}$ by a trace,
\begin{equation}
  \label{Weyl.hat.connection}
\hat{\omega}\indices{^{ab}}=\omega\indices{^{ab}}-h \delta^{ab}\,.
\end{equation}
Similarly, \eqref{connection.condition.2} yields
\begin{equation}
\label{spin.connection.2}
\hat{\omega}\indices{_\mu^{n}_{n}}=w_\mu\,.
\end{equation}
Again, the remaining component $\hat{\omega}\indices{_{\mu a}^n}$ of the Weyl spin connection is not determined from the specified geometric data.

Finally, the torsion of the Weyl spin connection may be written
\begin{subequations}
\begin{align}
T^{a}(\hat{\omega})&=de^{a}+ \omega^{ab} \wedge e^{b} -h \wedge e^{a}\,,\\
T^{n}(\hat{\omega})&=dn+w \wedge n+\hat{\omega}\indices{^{n}_{a}} \wedge e^{a}\,.
\end{align}
\end{subequations}

\paragraph{Induced Weyl connection at null infinity.}
We briefly return to the set-up and conventions of the main text to show explicitly that a Weyl connection for the unphysical metric is induced at null infinity. Recently, the role and appearance of a Weyl connection at the boundary of asymptotically AdS spacetimes has been similarly discussed \cite{Ciambelli:2019bzz}.

We take the conformal factor of the conformal compactification \eqref{eq:compactification} to be $\Omega=r^{-1}$.
The frame fields $\tilde{E}^{\sA}_\alpha$ of the unphysical manifold are related to the physical frame fields as follows,
\begin{equation}
  \label{eq:73}
  E^{\sf{a}}_\alpha=\Omega^{-1}\tilde{E}^{\sf{a}}_\alpha\,,\qquad E^{\sf{\hat{r}}}_\alpha=\Omega^{-2}\tilde{E}^{\sf{\hat{r}}}_\alpha\,,\qquad E^{\sf{\hat{u}}}_\alpha=\tilde{E}^{\sf{\hat{u}}}_\alpha\,.
\end{equation}
From the definition of the spin connection \eqref{eq:spincon}, we find
\begin{equation}
  \label{eq:77}
D_\alpha(\Omega^{-1}\tilde{E}^{\sf{a}}_\beta)=-\Omega^{-2}D_\alpha\Omega\tilde{E}^{\sf{a}}_\beta+\Omega^{-1}D_\alpha\tilde{E}^{\sf{a}}_\beta=-\Omega\indices{_\alpha^{\sf{a}}_{\sf{b}}} \Omega^{-1}\tilde{E}^{\sf{b}}_\beta+\Omega\indices{_\alpha^{\sf{a u}}} \tilde{E}^{\sf{\hat{u}}}_\beta+\Omega\indices{_\alpha^{\sf{a r}}} \tilde{E}^{\sf{\hat{r}}}_\beta\Omega^{-2}\,.
\end{equation}
Pulling back the above equation to $\scri^+$ and using the leading order of the spin connection  \eqref{eq:spinex}, one obtains
\begin{equation}
  \label{eq:78} D_i\tilde{E}^{\sf{a}}_j=-\omega\indices{_i^{\sf{a}}_{\sf{b}}}\tilde{E}^{\sf{b}}_j+\tilde{E}^{\sf{a}}_i h_j\,,
\end{equation}
or, using relation \eqref{Weyl.hat.connection} for the Weyl spin connection $\hat{\omega}$,
\begin{equation}
  \label{eq:55} D_i\tilde{E}^{\sf{a}}_j=-\hat{\omega}\indices{_i^{\sf{a}}_{\sf{b}}}\tilde{E}^{\sf{b}}_j\,.
\end{equation}
The bulk Levi-Civit\'a connection thus induces a Weyl connection when pulled back to $\scri^+$.

It may be useful to briefly relate this to Geroch's description of null infinity \cite{Geroch1977,Ashtekar:1981hw}. There, the freedom in performing Weyl rescalings is used to set $h=0$ in such a way that the induced connection at the boundary reduces to the more familiar metric compatible torsion-free covariant derivative. The undetermined piece in the connection then turns out to be the shear tensor describing, e.g., the radiative degrees of freedom. This field, however, is not part of the Chern--Simons connection $A$ as we only describe the vacuum sector.

\section{Newman--Unti gauge in the Newman--Penrose formalism}
\label{sec:comp-newm-penr}
Since the Newman--Unti gauge is usually discussed in terms of Newman--Penrose quantities, we provide a translation of the gauge and fall-off conditions of section \ref{sec:section2}. We will follow the conventions of \cite{Barnich:2016lyg} accounting for the different choice of signature in the metric which boils down to a change of sign in the Newman--Penrose (NP) coefficients.

Using the definition of the connection coefficients with respect to an orthonormal basis
\begin{equation}
  \label{eq:56}
\nabla_{\sA}E_{\sB}=\Omega\indices{_{\sA}^\sC_{\sB}}E_{\sC}\,.
\end{equation}
We decomposing the frame fields as
\begin{equation}
  \label{eq:68}
  l\equiv E_{\sf{\hat{r}}}=\partial_r\,, \qquad n\equiv E_{\sf{\hat{u}}}=\partial_u+U \partial_r+X^{i}\partial_i\,, \qquad m\equiv \frac{1}{\sqrt{2}}\left(E_{\sf{1}}+i E_{\sf{2}}\right)=\omega \partial_r+\xi^i\partial_i\,,
\end{equation}
with
\begin{subequations}
\begin{align}
\omega&=\frac{1}{\sqrt{2}}\left(E^r_{\sf{1}}+i E^r_{\sf{2}} \right)=-\frac{1}{\sqrt{2}}\left(E^i_{\sf{1}} E^{\sf{\hat{r}}}_i +i E^i_{\sf{2}} E^{\sf{\hat{r}}}_i \right)\,,\\
\xi^i&=\frac{1}{\sqrt{2}}\left(E^i_{\sf{1}}+i E^i_{\sf{2}} \right)\,,\\
X^i&=-E^{\sa}_u E_{\sa}^i\,.
\end{align}
\end{subequations}
Thus, the relation between coordinate and frame components is given by
\begin{subequations}
\begin{align}
  \label{eq:69}
  \Omega_{ \sf{\hat{r}AB}}&=\Omega_{r\sf{AB}}\,\\
  \Omega_{\sf{\hat{u}} \sf{AB}}&=\Omega_{u\sf{AB}}+U \Omega_{r\sf{AB}}+X^i\Omega_{i\sf{AB}}\,\\
  \Omega_{m \sf{AB}}&=\omega \Omega_{r\sf{AB}}+\xi^i\Omega_{i\sf{AB}}\,.
\end{align}
\end{subequations}
The complex NP coefficients which will play a role in the following are
\begin{subequations}
\begin{align}
  \label{eq:70}
  \kappa&\equiv \Omega_{\sf{\hat{r}\hat{r}m}}=-l^{\alpha}m^\beta \nabla_\alpha l_\beta=\Omega_{r\sf{\hat{r}m}}=\frac{1}{\sqrt{2}}\left(\Omega_{r\sf{\hat{r}1}}+i \Omega_{r\sf{\hat{r}2}}\right)\,,\\
  \pi&\equiv -\Omega_{\sf{\hat{r}\sf{\hat{u}}\bar{m}}}= l^\alpha\bar{m}^\beta\nabla_\alpha n_\beta=-\frac{1}{\sqrt{2}}\left(\Omega_{r\sf{\hat{u}1}}-i \Omega_{r\sf{\hat{u}2}}\right)\,,\\
  \epsilon&\equiv  \frac{1}{2}\left(\Omega_{\sf{\hat{r}\hat{r}\hat{u}}}-\Omega_{\sf{\hat{r}m\bar{m}}}\right)=\frac{1}{2}l^\alpha\left(\bar{m}^\beta\nabla_\alpha m_\beta-n^\beta\nabla_\alpha l_\beta\right)=\frac{1}{2}\left(- \Omega_{r\sf{12}}-\Omega_{r\sf{\hat{u}}\hat{r}}\right)\,,\\
  \rho&\equiv  \Omega_{\sf{\bar{m}\hat{r}m}}=-\bar{m}^\alpha m^\beta\nabla_\alpha l_\beta=\bar{\omega}\kappa +\frac{1}{\sqrt{2}}\xi^i\left(\Omega_{i\sf{\hat{r}1}}+i \Omega_{i\sf{\hat{r}2}}\right)\,,\\
  \tau&\equiv \Omega_{\sf{\hat{u}\hat{r}m}}=-n^\alpha m^\beta\nabla_\alpha l_\beta=\frac{1}{\sqrt{2}}\left(\Omega_{u\sf{\hat{r}1}}+i \Omega_{u\sf{\hat{r}2}}\right)+U\kappa+\frac{X^i}{\sqrt{2}}\left(\Omega_{i\sf{\hat{r}1}}+i \Omega_{i\sf{\hat{r}2}}\right)\,,\\
  \bar{\alpha}+\beta&\equiv\Omega_{\sf{m\hat{r}\hat{u}}}= m^\alpha n^\beta\nabla_\alpha l_\beta=\omega\Omega_{r\sf{\hat{r}\hat{u}}}+\xi^i\Omega_{i\sf{\hat{r}\hat{u}}}\,.
\end{align}
\end{subequations}
The Newman--Unti gauge-fixing conditions are given by \cite{Newman:1962cia,Newman:1961qr,Barnich:2019vzx}
\begin{equation}
  \label{eq:NUgauge}
  \kappa=\epsilon=\pi=0\,, \qquad \rho=\bar{\rho}\,, \qquad \tau=\bar{\alpha}+\beta\,.
\end{equation}
Imposing the six real conditions of the first equation yields
\begin{equation}
  \label{eq:72}
  \Omega\indices{_r^{\sf{AB}}}=0\,.
\end{equation}
From the definition of $\xi^i$, we find
\begin{equation}
  \label{eq:74}
\rho=\frac{1}{2}\left(E_{\sf{a}}^i\Omega\indices{_i^{\sf{a}\sf{\hat{u}}}}+i \epsilon\indices{^{\sf{a}}_{\sf{b}}}E_{\sf{a}}^i\Omega\indices{_i^{\sf{b}\sf{\hat{u}}}}\right)\,,
\end{equation}
so that the gauge-fixing condition \eqref{eq:NUgauge} forces the second term to vanish,
\begin{equation}
  \label{eq:75}
\epsilon\indices{^{\sf{a}}_{\sf{b}}}E_{\sf{a}}^i\Omega\indices{_i^{\sf{b}\sf{\hat{u}}}}=0\,.
\end{equation}
Finally, the last condition in \eqref{eq:NUgauge} yields
\begin{equation}
  \label{eq:71}
  \Omega_{u\sf{\hat{r}a}}+X^i\Omega_{i\sf{\hat{r}a}}-\Omega_{i\sf{\hat{r}\hat{u}}}E_{\sa}^i=0\,.
\end{equation}
Together with \eqref{eq:75}, this is equivalent to the constraint $T^{\sf{\hat{u}}}_{ui}=0$. In summary, we find that the gauge described in section~\ref{sec:newman-unti-solution} coincides with the Newman--Unti gauge.

In addition to the above gauge conditions, the Newman--Unti solution space also implements certain fall-off conditions. These are given by
\begin{align}
  \label{eq:76}
  X^i=O(r^{-1})\,, \quad \rho=-\frac{1}{r}+O(r^{-3})\,, \quad \tau=O(r^{-2})\,, \quad \Psi_0=\Psi_0^0\ r^{-5}+ O(r^{-6})\,.
\end{align}
Using the fields asymptotic expansion \eqref{eq:framefall} and \eqref{eq:spinex}, we find that the first condition is satisfied. The statement that the $r^{-2}$ term is absent in the expansion of $\rho$ leads to the requirement that the trace of $C_{ij}$ vanishes. An explicit calculation shows that this last condition is satisfied as well. However, the fall-off behavior of the spin coefficient $\tau$ is $O(r^{-1})$, weaker than required by the above condition due to the presence of a non-zero $h_i$.

In summary, the gauge used in the main text coincides with the Newman--Unti gauge apart from the slower fall-off in $\tau$ when $h_i\neq 0$. The latter quantity is pure gauge however, and may be set to zero by a null rotation around $E^{\sf{\hat{u}}}$ \cite{Newman:1962cia}.

\section{Extension to the gauged Poincaré algebra}
\label{sec:extendedalgebra}
We showed in section~\ref{sec:radiative-vacua} that the first layer of asymptotic equations, found by expanding the gravitational field equations near $\scri^+$ and by imposing absence of gravitational radiation, coincides with the equations of motion of a $\mathfrak{so}(3,1) = \mathfrak{conf}(2)$ Chern--Simons theory. Here, we extend this analysis and show that this first layer, \emph{together} with the second layer of bulk equations described in section~\ref{sec:subleading}, may be recast as vanishing of the field-strength of a gauge connection $A$ for the Poincaré or three-dimensional conformal Carrollian algebra $\mathfrak{iso}(3,1)=\mathfrak{ccar}(3)$,
\begin{equation}
\label{eq:EOM-ccar}
F\left[A\right]\equiv \dd A+A \wedge A=0.
\end{equation}
To be more specific, we consider the gauge field
\begin{equation}
  \label{eq:96}
  A=n \Pt_0+e^{\sf{a}}\Pt_{\sf{a}}+\omega \Jt+h \Ht+b^{\sf{a}}\Bt_{\sf{a}}+ e^{(0)\sf{a}}\Ct_{\sf{a}}+h^{(0)} \Bt\,,
\end{equation}
where generators satisfy the algebra
\begin{subequations}
\begin{align}
  \label{eq:95}
  [\Bt_{\sf{a}},\Ct_{\sf{b}}]&=\Bt\delta_{\sf{ab}}\,,\qquad &&[\Ht,\Bt]=\Bt\qquad &&[\Pt_{\sf{a}},\Ct_{\sf{a}}]=\delta_{\sf{ab}}\Pt_0\,,\\
  [\Pt_0,\Bt_{\sf{a}}]&=-\Ct_{\sf{a}}\,,\qquad &&[\Pt_{\sf{a}},\Bt]=\Ct_{\sf{a}}\,,\qquad &&[\Ht,\Pt_0]=-\Pt_0\\
  [\Bt_{\sf{a}},\Pt_{\sf{b}}]&=\Ht\delta\indices{_{\sf{ab}}}-\Jt\epsilon\indices{_{\sf{ab}}}\,,\qquad &&[\Ht,\Pt_{\sf{a}}]=-\Pt_{\sf{a}}\,,\qquad&&[\Ht,\Bt^{\sf{a}}]=\Bt^{\sf{a}}\\  [\Jt,\Bt^{\sf{a}}]&=\epsilon\indices{^{\sf{ab}}}\Bt^{\sf{b}}\,\qquad &&[\Jt,\Pt_{\sf{a}}]=\epsilon\indices{_{\sf{ab}}}\Pt_{\sf{b}} \qquad && [\Jt,\Ct^{\sf{a}}]=\epsilon\indices{^{\sf{ab}}}\Ct^{\sf{b}}\,.
\end{align}
\end{subequations}
Here, $n$ is identified with the `missing' boundary frame field giving the time direction,
\begin{equation}
  \label{eq:98}
  n=\dd u\,.
\end{equation}
Korovin \cite{Korovin:2017xqu} already pointed out the appearance of a gauged $\mathfrak{ccar}(3)$ algebra governing the onshell residual symmetry transformations of the asymptotic fields $e^{\sa}, h, \omega, b^{\sa},$ $\tau, e^{(0)\sa}, h^{(0)}$. In his analysis, Korovin used weaker gauge and boundary conditions that allowed him to keep full covariance with respect to boundary coordinates and boundary frame rotations. In the present analysis, a choice of time coordinate and further partial gauge fixing have been made according to the standard Newman--Unti framework \cite{Newman:1962cia}.

Computing the field-strength components, we recover \eqref{eq:eomCS} together with
\begin{subequations}
    \label{eq:exteom}
\begin{align}
  F(\Pt_0)&=\dd n-h\wedge n+e^{\sf{a}}\wedge e^{(0)}_{\sf{a}}\,,\\
  F(\Bt)&=\dd h^{(0)}+h \wedge h^{(0)}+b^{\sf{a}}\wedge e^{(0)}_{\sf{a}}\,,\\
  \label{eq:dc-app}
  F(\Ct)^{\sf{a}}&=\dd e^{(0)\sf{a}}-\epsilon^{\sf{ab}}\omega\wedge e^{(0)\sf{b}}-n\wedge b^{\sf{a}}-h^{(0)}\wedge e^{\sf{a}}\,.
\end{align}
\end{subequations}
Hence, the subleading bulk equation \eqref{eq:dc} derived in section~\ref{sec:subleading} is nothing but the vanishing of $F(\Ct)^{\sf{a}}$. The first two equations, although they do not appear as field equations from our bulk analysis, are shown to hold identically thanks to the choice of boundary conditions made in section~\ref{sec:newman-unti-solution}. It is very likely, although we have not checked it explicitly, that these would (re-)appear as bulk field equations if these boundary conditions were relaxed.  

As also shown in \cite{Korovin:2017xqu}, the onshell residual symmetry transformations of these asymptotic fields are given by \eqref{onshell.sym} together with 
\begin{align}
  \label{eq:105}
  \delta n&=\dd \pi+\eta n-\pi h-\pi_{\sf{a}}e^{(0)\sa}+\gamma_{\sf{a}}e^{\sf{a}}\,,\\
  \delta h^{(0)}&=\dd \zeta+\zeta h -\eta h^{(0)}-\lambda_{\sf{a}}e^{(0)\sf{a}}+\gamma_{\sf{a}}b^{\sf{a}}\,,\\
   \delta e^{(0)\sf{a}}&=\dd \gamma^{\sf{a}}+\zeta e^{\sf{a}}-\pi^{\sf{a}}h^{(0)}+\pi b^{\sf{a}}- \lambda^{\sf{a}}n+\epsilon\indices{^{\sa}_{\sf{b}}}(\lambda e^{(0)\sf{b}}-\gamma^{\sf{b}}\omega)\,,
\end{align}
where new gauge parameters have been introduced,
\begin{subequations}
\label{eq:onshell-transf-extended}
\begin{align}
\pi&\equiv \xi^u\,,\\
\zeta &\equiv -h^i \partial_i \xi^u+\xi^u h_u^{(0)}+Y^i h_i^{(0)}+\alpha \,,\\
\gamma^{\sa}&\equiv-\partial_{\sa} \xi^u+\xi^u h^{\sa}+Y^i e_i^{(0)\sa}\,.
\end{align}
\end{subequations}
In the present context, $n$ has been gauged-fixed through \eqref{eq:98} such that $\delta n$ actually vanishes.
The onshell transformations \eqref{eq:onshell-transf-extended} coincide with (a gauged-fixed version of) the gauge transformation
\begin{equation}
  \label{eq:97}
\delta_\Lambda A=\dd \Lambda+[A,\Lambda], \qquad
 \Lambda=\pi \Pt_0+\pi^{\sa}\Pt_{\sa}+\lambda\Jt+\eta \Ht+\lambda^{\sa}\Bt_{\sa}+\gamma^{\sf{a}}\Ct_{\sf{a}}+\zeta \Bt\,.
\end{equation}

At the level of equations of motion and onshell symmetry transformations, we thus exhibit a gauging of the $\mathfrak{iso}(3,1)=\mathfrak{ccar}(3)$ algebra. However, it is well-known that there is no Chern--Simons action whose equations of motion would yield \eqref{eq:EOM-ccar}, in contrast to the situation encountered for the gauged $\mathfrak{so}(3,1)$ algebra described in section~\ref{sec:CStheory}. This is due to a lack of nondegenerate bilinear form on the Poincaré algebra. As a curiosity, we mention that an action reproducing the subset of equations of motion \eqref{eq:exteom} may be found,
\begin{equation}
S=\int  2n \wedge h^{(0)} \wedge h-2 n \wedge \dd h^{(0)} + 2n \wedge e^{(0)\sa} \wedge b_{\sa}+ e^{(0)\sa} \wedge \dd e^{(0)}_{\sa}- 2 e_{\sa} \wedge e^{(0)\sa} \wedge h^{(0)}- \omega_{\sf{ab}} \wedge e^{(0)\sa} \wedge e^{(0)\sf{b}}.
\end{equation}
Here $n, h^{(0)}, e^{(0)\sa}$ are the only dynamical fields, while $e^{\sa}, \omega, h, b^{\sa}$ are considered part of a fixed background structure. This action is invariant under the full set of $\mathfrak{iso}(3,1)=\mathfrak{ccar}(3)$ gauge transformations transforming both dynamical and background fields. 

\bibliography{bibl}
\bibliographystyle{JHEP}
\end{document}